\documentclass[aps,pra,twocolumn,a4paper,reprint,superscriptaddress,nofootinbib,floatfix]{revtex4-1} 
\pdfoutput=1
\usepackage{amsmath}
\usepackage{amssymb}
\usepackage{dsfont}
\usepackage{verbatim}
\usepackage{graphicx}
\usepackage{tabularx}
\usepackage{xcolor}
\usepackage{mathtools}
\usepackage{bbold}
\usepackage{blkarray}
\usepackage{appendix}
\usepackage[shortlabels]{enumitem}
\usepackage{latexsym}
\usepackage{array}
\usepackage{colortbl}
\usepackage{bm} % for bold math symbols
\usepackage{amsthm}
\usepackage{siunitx}

\usepackage{thmtools}

\newcommand{\Order}[2][1]{\mathcal{O}\left(#2\right)}

\usepackage{accents}

\DeclareSIUnit{\gauss}{G}
\DeclareSIUnit{\phonons}{phonons}

\newcommand{\bra}[1]{\mathinner{\langle #1|}}
\newcommand{\ket}[1]{\mathinner{|#1\rangle}}

\newcommand{\ketbra}[2]{\mathinner{|#1\rangle\langle#2|}}

\usepackage{dsfont} % for double stroke letters
\usepackage{hyperref}
\usepackage{nicefrac}
\hypersetup{colorlinks=true,        % false: boxed links; true: colored links
    linkcolor=red,         % color of internal links
    citecolor=blue,        % color of links to bibliography
    filecolor=black,      % color of file links
    urlcolor=blue }

\newcommand{\uibkexp}{\affiliation{Universit\"at Innsbruck, Institut f\"ur Experimentalphysik, Technikerstraße 25, Innsbruck, Austria}}
\newcommand{\rwth}{\affiliation{Institute for Quantum Information, RWTH Aachen University, D-52056 Aachen, Germany}}
\newcommand{\fzj}{\affiliation{Institute for Theoretical Nanoelectronics (PGI-2), Forschungszentrum J\"ulich, D-52425 J\"ulich, Germany}}

\begin{document}

\title{Demonstration of Logical Gates on a Fault-Tolerantly Encoded Logical Qutrit}
\author{Peter Tirler$^*$}
\uibkexp
\author{Irina Heinz$^*$}
\fzj
\rwth
\author{Lukas Gerster}
\uibkexp
\author{Manuel John}
\uibkexp
\author{Keshav Pareek}
\uibkexp
\author{Tim Gollerthan}
\uibkexp
\author{Michael Meth}
\uibkexp
\author{Markus Müller}
\fzj
\rwth
\author{Martin Ringbauer}
\uibkexp

\begingroup
\renewcommand{\thefootnote}{}
\footnotetext{* These authors contributed equally.}
\endgroup

\begin{abstract}

Quantum computing is one of the most anticipated technologies of the current decade. 
Yet, most quantum processors today still use binary information encoding, while the underlying hardware almost always supports higher-dimensional encoding in so-called qudits. The use of qudits promises larger Hilbert spaces and more efficient quantum information processing. 
Like qubits, however, qudits are susceptible to errors, and quantum error correction in conjunction with fault-tolerant gate operations is expected to play a key role in achieving practical quantum advantage. 
Here, we develop and experimentally realize, on a trapped-ion quantum processor, a complete qudit quantum-error correction code, generalizing the qubit Steane code to qudits of local dimension $d = 3$. We fault-tolerantly encode the logical qutrit in the basis state $\ket{0}_{\rm L}$ with a fidelity of $\text{F}=0.947(7)$, and perform sequences of up to 81 logical single-qutrit Clifford gates with an average fidelity of $\text{F}=0.9923(4)$. We further demonstrate fault-tolerant logical magic-state preparation, a key ingredient for universal quantum computation with encoded qutrits, with a fidelity of $\text{F}=0.96(3)$. This fault-tolerant demonstration of a qutrit QEC code with logical gate operations paves the way towards error-corrected qudit quantum computing. 
\end{abstract}

\maketitle

% --------------------------
% INTRODUCTION
% --------------------------
\section{Introduction}
\label{sec:introduction}
% --------------------------
Modern quantum computers predominantly follow the paradigms established in classical computing. Notably, quantum information is encoded in \textit{qubits}, which are defined by two computational basis states labeled by $\ket{0}$ and $\ket{1}$, respectively. Most hardware platforms, on the other hand, naturally offer a larger manifold of states that can be used to encode quantum information. Exploiting these states enables computation with higher-dimensional quantum systems, known as \textit{qudits}, and recent progress in experimental control has greatly increased interest in qudit-based quantum computing~\cite{wang2018, Ringbauer2022, Yu2025, Morvan2021, Chi2022, Low2026}. In a qudit, quantum information is stored not in two, but in $d$ quantum states $\ket{0},\dots, \ket{d-1}$ per information carrier. This provides a more adequate use of the available resources, e.g., the multiple energy levels of an ion. Qudits then enable the storage and manipulation of information beyond the capabilities of purely qubit-based architectures, allowing for a more efficient implementation of certain quantum algorithms, such as for quantum simulation and quantum phase estimation~\cite{Wang2020, Meth2025, Goss2026, john2026, Asaduzzaman2026}.

Like all quantum processors, qudit architectures are susceptible to errors arising from unwanted interactions with the environment and imperfect gate calibration. In qudits, this is particularly relevant, since larger local Hilbert spaces can lead to more complex noise behaviors. 
This inherent fragility of quantum information has given rise to an extensive study of quantum error correction (QEC) protocols %~\cite{Shor1995, Steane1996, Kitaev1998}, 
and the demonstration of a variety of quantum codes on various hardware modalities~\cite{Abobeih2022, Krinner2022, Brock2025, Bluvstein2026, Acharya2025, Paetznick2026}. Among these, trapped ions constitute a highly promising platform to realize scalable quantum computing~\cite{Egan2021, Erhard2021, Postler2022, lavasani2026, Rojkov2024, tham2026breakevendemonstrationquantumlowdensity}. Despite impressive progress, however, most QEC codes demonstrated to date have been limited to protecting and processing quantum information encoded in binary systems. Qudit QEC, on the other hand, remains much less developed.
Recently, one physical qudit in a single trapped ion has been used to reduce the idling error of one locally encoded qubit \cite{DeBry2026}, and Gottesmann-Kitaev-Preskill~\cite{Gottesman2001} codes have demonstrated the detection and correction of idling noise for a qudit encoded in a single harmonic oscillator mode~\cite{Brock2025}.

For implementing complete, scalable QEC codes with fault-tolerant quantum operations, however, the most widely used approach involves stabilizer codes, where quantum information is redundantly encoded in an entangled state across multiple physical systems, and errors can be detected through measurements of stabilizer operators without revealing the encoded logical qudit state.
It has been shown that many such stabilizer QEC codes, including color codes, can be generalized to higher-dimensional systems~\cite{Grassl2003, PhysRevA.87.062338, PhysRevA.92.022312, Spencer_2026}, and that qudit codes can enable more efficient distillation of magic states~\cite{PhysRevX.2.041021, PhysRevLett.113.230501}. The latter are a key resource for the implementation of universal gate operations, which prevent errors from propagating uncontrollably, commonly referred to as fault-tolerant (FT) quantum computation. However, scalable qudit QEC, as well as FT gate operations on logical qudits, have remained elusive. 

Here we develop quantum protocols to realize a qutrit ($d=3$) generalization of the binary [[7,1,3]] Steane code~\cite{Steane1996}, see Fig.~\ref{fig:0_trap_schematic}. It encodes one logical qutrit in seven physical qutrits with a code distance of 3, allowing for the correction of an arbitrary single-qutrit error in the system. To this end, we use ten physical qutrits in a trapped-ion quantum processor, seven of which are used to encode one logical qutrit, and three to act as ancilla qutrits for stabilizer readout. Inspired by existing qubit implementations, we develop and implement protocols for FT preparation of logical basis states, as well as FT logical Clifford gates on the encoded qutrit using transversal operations. Finally, we achieve the FT preparation of a logical qutrit magic state as a key resource to enable FT $T$ gates by magic state injection.  

\section{The Qutrit Steane Code}
Qutrit operations can be described using the generalized Pauli operators, $X$ and $Z$, acting as clock and shift operators in the Hilbert space. The qutrit Clifford group is then generated by the Hadamard gate $H$, the phase gate $S$, and the two-qutrit $\mathrm{CSUM}$ gate, the qutrit analogue of the qubit $\mathrm{CNOT}$ gate. Their matrix representations are given in Section~\ref{sec:methods}.

\begin{figure}[t]
    \centering
    \includegraphics[width=1.0\columnwidth]{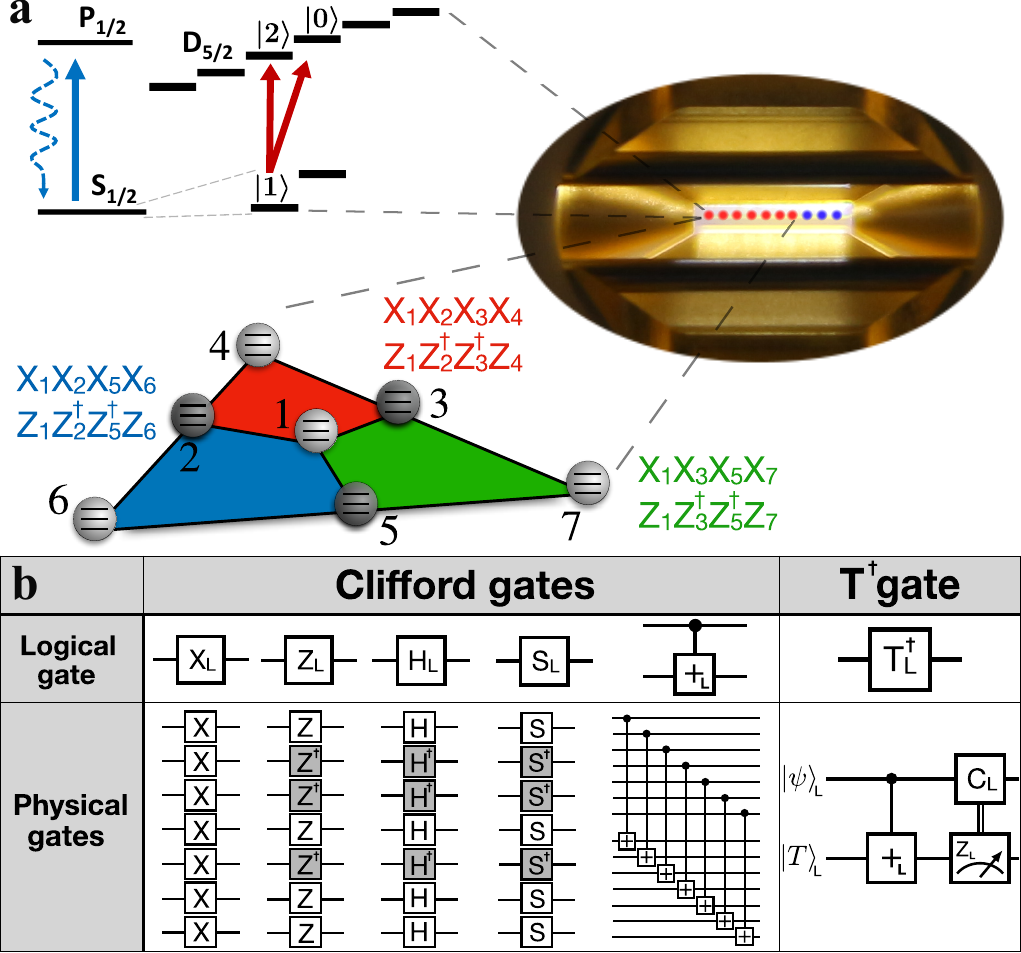}
    \caption{\textbf{a}: The top right shows the ion trap used for the experiment, including an illustration of the seven data qutrits (red) and three ancilla qutrits (blue). To the left, the energy level diagram of $^{40}\text{Ca}^+$ is shown, restricted to the most important transitions. The three levels encoding the qutrit information are labeled by $\ket{0}$, $\ket{1}$, and $\ket{2}$. Red arrows represent transitions used for driving physical gates by coupling the qutrit levels with narrow-band laser light. The blue arrows correspond to an auxiliary transition exhibiting fast spontaneous decay, indicated by the curly arrow, which is used for laser cooling of the ion chain and readout of the individual ions' states. The schematic below shows the Steane code with its stabilizers, with plaquette operators highlighted in red, blue, and green. 
    \textbf{b}: Summary of logical gates in the qutrit Steane code and their implementation on the physical qutrits. Clifford gates are transversal, and thereby inherently FT, whereas $T^{\dagger}$ gates can be implemented using magic-state injection. At the qutrits shaded in dark gray, the $Z$, $H$, and $S$-operators are inverted.}
    \label{fig:0_trap_schematic}
\end{figure}

The qutrit Steane code, encoding one logical qutrit, is defined by three $X$-type and three $Z$-type stabilizer operators, each acting on four out of a total of seven physical qutrits. The code space is defined as the common +1 eigenspace of the six mutually commuting stabilizers. The code can be visualized as a triangle consisting of three plaquettes, with two stabilizers assigned to each plaquette, as shown in Fig.~\ref{fig:0_trap_schematic}a. The three plaquettes, corresponding to different regions of the triangle, are assigned different colors, and the code represents the smallest instance of the scalable family of two-dimensional qutrit color codes \cite{Kubica_2015, PhysRevA.92.022312}. Note that the $Z$-stabilizers are composed of both physical $Z$- and $Z^{\dagger{}}$ operations, the latter at positions where two plaquettes share one qutrit, depicted in dark gray.

In this code, the logical Pauli operators $X_{\rm L}$ and $Z_{\rm L}$, as well as the logical Clifford group generators $H_{\rm L}$, $S_{\rm L}$, and $\mathrm{CSUM}_{\rm L}$ are transversal, meaning that any of these logical gates can be implemented as a parallel execution of physical gates on all data qutrits, as shown in Fig.~\ref{fig:0_trap_schematic}b. The exact decomposition of these operators into physical operations and their action is discussed in detail in Section~\ref{sec:methods}.This transversality ensures fault-tolerance of these gate operations by construction, since potential single physical errors during the execution remain localized and lead to at most a single, and thus correctable, error per logical qutrit block.

In addition to Clifford gates, one non-Clifford gate, such as the qutrit $T$ gate, is needed for a universal gate set. The Eastin-Knill theorem~\cite{Eastin2009} states that a universal set of gates cannot be implemented transversally in a QEC code. Since all Clifford gates are transversal in the qutrit Steane code, the $T$ gate cannot be transversal in the same code, and its FT implementation requires different methods such as magic-state injection~\cite{Bravyi2005} using a logical resource state, depicted in the right column of Fig.~\ref{fig:0_trap_schematic}b. The main challenge lies in the fault-tolerant preparation of the logical qutrit magic state, which can subsequently be injected using a transversal CSUM gate.

The quantum processor used in the experiment consists of ten $^{40}\text{Ca}^+$-ions in a macroscopic linear Paul trap~\cite{Ringbauer2022}, see Section~\ref{sec:methods} for details. Fig.~\ref{fig:0_trap_schematic}a shows a picture of the trap with a schematic of the linear ion chain and an overview of the $^{40}\text{Ca}^+$ electronic energy structure, highlighting the levels used to encode a physical qutrit.

\section{Logical state encoding}

As a first step, we fault-tolerantly prepare the state $\ket{0}_{\rm L}$ of the Steane code, defined as the eigenstate of the logical $Z_{\rm L}$ operator with eigenvalue +1. To this end, we start with a non-FT unitary encoding circuit acting on the seven qutrits. This is followed by measurements of stabilizers and the logical Pauli $Z_{\rm L}$ operator to detect all higher-weight errors, which affect two or more physical qutrits and may therefore be uncorrectable by the code~\cite{Goto2016}. If any of these measurements returns a non-trivial outcome, a possibly uncorrectable error has occurred and the preparation run is discarded. This results in a repeat-until-success protocol that fault-tolerantly prepares $\ket{0}_{\rm L}$, such that an arbitrary single fault can result at most in a single, and thus correctable error.

\begin{figure*}[t]
    \centering
    \includegraphics[width=0.96\textwidth]{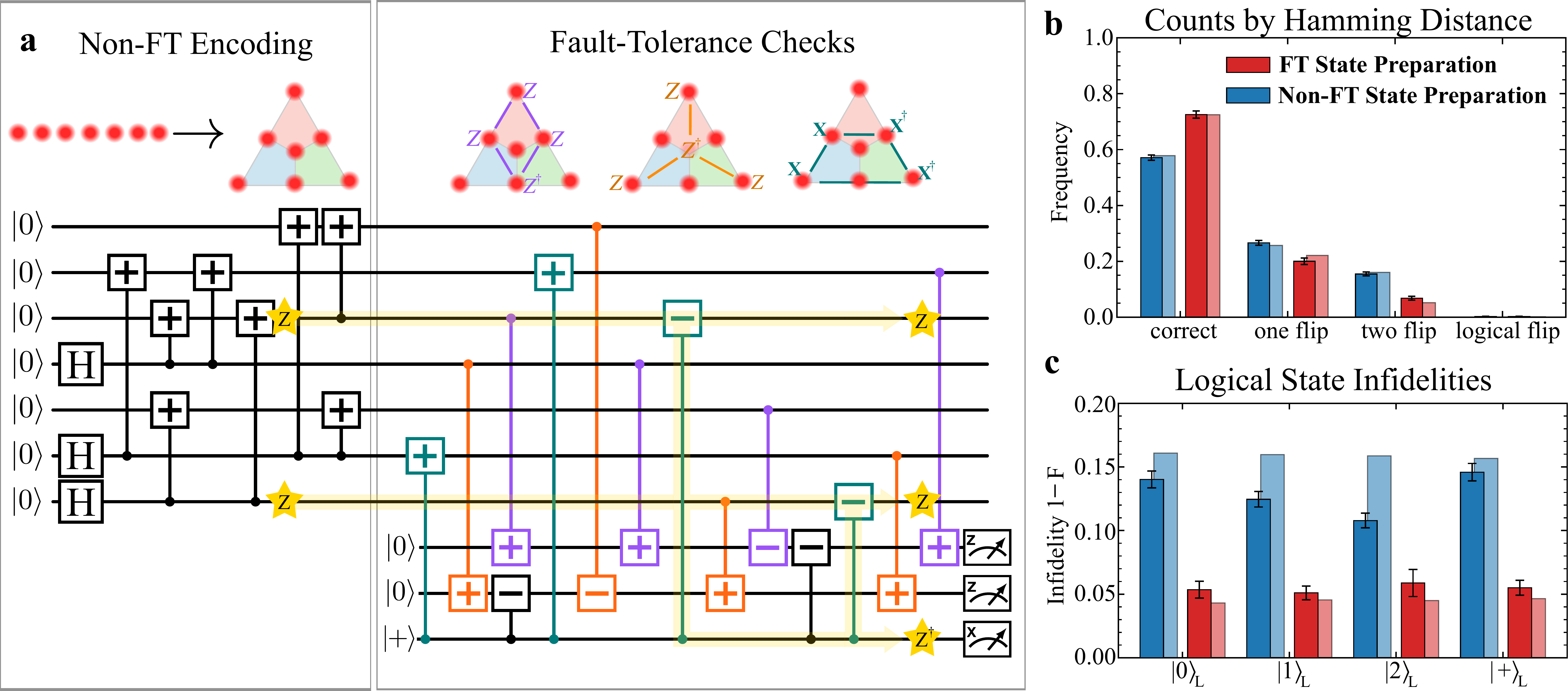}
    \caption{ \textbf{a}: Fault-tolerant encoding protocol for the logical $\ket{0}_{\rm L}$ state of the qutrit Steane code. First, seven qutrits are non-fault-tolerantly encoded into the $\ket{0}_{\rm L}$-state of the Steane code. This is followed by a flag-fault-tolerant measurement of the operators $Z_2Z_3Z_4Z_5^{\dagger}$, $Z_1^{\dagger}Z_4Z_6Z_7$, and $X_2X_3^{\dagger}X_6X_7^{\dagger}$. The yellow stars correspond to an exemplary weight-two $Z$-type fault, which would correspond to a non-correctable error if left undetected, but is captured by the $X_2X_3^{\dagger}X_6X_7^{\dagger}$ stabilizer measurement. CSUM and CSUM$^{\dagger}$ gates are represented with controlled-plus and controlled-minus symbols, respectively. \textbf{b}:~Relative occurrences of logical output states by Hamming distance. Measurement outcomes in the $Z$-basis are grouped by how many trit-flip errors they differ from the target $\ket{0}_{\rm L}$-state. The blue bars correspond to a direct state preparation using only the first part of the circuit in \textbf{a}, while the red bars correspond to the fault-tolerant encoding using three ancillas. Error bars represent one standard deviation of stochastic uncertainty (shot noise). Dark color bars correspond to experimental data and lighter colors are the corresponding simulated values. \textbf{c}: Total state preparation infidelities for various logical states, prepared fault-tolerantly and non-fault-tolerantly. This includes decoding of the syndrome information after the measurement.}
    \label{fig:all}
\end{figure*}

In fact, for the non-FT encoding shown in Fig.~\ref{fig:all}a, it suffices to measure two representatives of the logical $Z_{\rm L}$ operator, $Z_2Z_3Z_4Z_5^{\dagger}$ and $Z_1^{\dagger}Z_4Z_6Z_7$, and one $X$ stabilizer representative $X_2X_3^{\dagger}X_6X_7^{\dagger}$.
The quantum circuit of this state-preparation scheme, including the propagation and detection of one example error, is depicted in Fig.~\ref{fig:all}a. The error $Z_3 Z_7$ arising from a CSUM gate in the encoding circuit, acts on qutrits 3 and 7. It cannot be reduced to an error of lower weight error, i.e.~one acting on fewer physical qutrits, by multiplication with stabilizers or the logical operator $Z_{\rm L}$, and therefore represents a genuine weight-two fault. In fact, a $Z$-type error arising from this CSUM gate corresponds to an actual weight-two error for $Z_3Z_7$ or $Z_3^{\dagger}Z_7^{\dagger}$. In contrast, $Z_3^{\dagger}Z_7$ and $Z_3Z_7^{\dagger}$ are effective weight-one errors, as they differ by only a single-qutrit error from the logical operators $Z_L$ and $Z_L^{\dagger}$, represented by $Z_4Z_3^{\dagger}Z_7$ and $Z_4^{\dagger}Z_3Z_7^{\dagger}$, respectively, which stabilize the encoded state. This is a subtle but important difference from the qubit case. Capturing these errors requires measuring at least one $X$-type stabilizer.

Measuring a four-body operator like  $Z_2Z_3Z_4Z_5^{\dagger}$ non-destructively requires entangling an ancilla qutrit with all four qutrits on which the operator acts. This introduces another error mechanism, where an error on this ancilla can spread onto the logical state. A fault-tolerant readout of the operators $Z_2Z_3Z_4Z_5^{\dagger}$, $Z_1^{\dagger}Z_4Z_6Z_7$, and $X_2X_3^{\dagger}X_6X_7^{\dagger}$ can nonetheless be achieved by introducing a flag qutrit in addition to the ancilla for each measurement~\cite{PhysRevLett.121.050502}. A nontrivial measurement outcome of the flag qutrit indicates the potential presence of a harmful error that cannot be corrected by the code, e.g., if it occurs on the ancilla and propagates onto two qutrits of the encoded state. By postselecting on trivial outcomes for both the ancilla and the flag qutrit, we suppress these uncorrectable errors. Notably, following the approach proposed in~\cite{Reichardt_2021}, these measurements can be performed in parallel, with the ancilla qutrits associated with different operators acting mutually as flags. This eliminates the need for additional dedicated flag qutrits, allowing the encoding to be implemented using only ten qutrits in total. Fig.~\ref{fig:all}a shows the resulting repeat-until-success, fault-tolerant encoding protocol.

As a reference measurement, we first prepare the logical $\ket{0}_{\rm L}$ state non-fault-tolerantly. In a second experiment, we execute the entire circuit, including the parallel fault-tolerance checks. A measurement run is discarded whenever at least one ancilla qutrit is found in a state other than its initial state. The result of each measurement run either belongs to one of the states in the code space, or to a state outside of it, being either one or two physical errors away from the targeted $\ket{0}_{\rm L}$ state. Since the final measurement is performed in the physical $Z$ basis, we count the number of $X$-flips, or trit-flips, that a measurement result differs from the target state, known as the Hamming distance. Figure~\ref{fig:all}b shows the experimentally observed statistics of measurement outcomes in the $Z$-basis grouped by the number of trit-flips relative to the state $\ket{0}_{\rm L}$. A logical flip results in a state that lies in the codespace but is different from $\ket{0}_{\rm L}$, which requires flips on at least three physical qutrits to happen. In our experiment, the flag-fault-tolerance check accepts 639 out of 3100 measurement runs where no ancilla qutrit raises an error, while reducing the relative likelihood of potentially uncorrectable weight-two errors from 15.7(8)\% to 7.0(8)\%. We compare this to a stabilizer-tableau simulation~\cite{brandl2026} of the qutrit circuit using a multi-parameter Clifford noise model, which agrees within the shot-noise uncertainties. The noise model and simulations are discussed in more detail in Appendix~\ref{sec:error-model}.

After measuring the logical qutrit, we can perform a virtual round of (ideal) error correction in postprocessing. This assigns all states associated with one physical flip to the corresponding closest state within the code space. Notably, unlike the qubit-based Steane code, in the qutrit case this is not enough to fully decode any possible syndrome, since there exist error syndromes of weight two relative to any of the logical states. We therefore construct a decoder which assigns all weight-one syndromes and ambiguous weight-two syndromes to the error they most likely correspond to, and then corrects them accordingly. For more details, see the Appendix~\ref{sec:decoder}.

Decoding all outcomes to one of the logical states allows us to assign logical quantum state fidelities. For the $\ket{0}_{\rm L}$-state, this results in a fidelity of F = 0.860(7) for the non-FT state preparation, and a FT state-preparation fidelity of F = 0.947(7), highlighting the performance benefit of the significantly more complex FT protocol over the non-FT circuit. Note that the latter value is larger than the sum of the no-error and one-error likelihoods, which add up to 0.926(8), since the constructed decoder can also assign some outcomes with weight-two errors to the correct logical state. Furthermore, the state fidelities of all computational basis states as well as one $X$-basis state are extracted via linear inversion logical state tomography (see Appendix~\ref{sec:tomogrphy} for more details), all of which can be prepared with a fidelity within [0.94, 0.95], as shown in Figure~\ref{fig:all}c. For comparison, when the same states are prepared without the fault-tolerance checks, the fidelity lies in  [0.85, 0.90]. States different from $\ket{0}_{\rm L}$ are prepared by applying transversal logical operations, as discussed in the next section, following the respective non-FT and FT preparation circuits for $\ket{0}_{\rm L}$.

\begin{figure}[ht]
    \centering        \includegraphics[width=\linewidth]{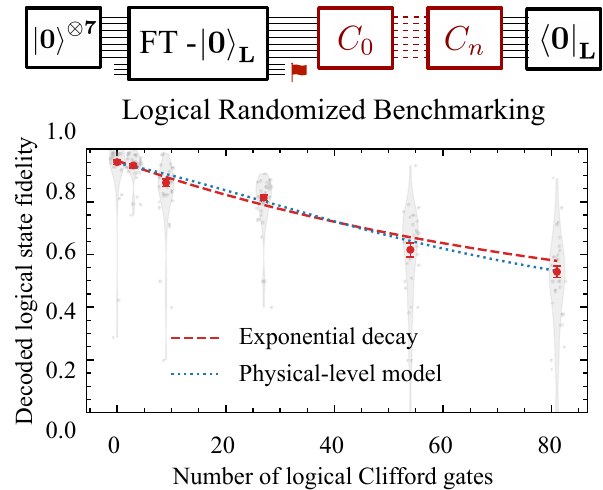}
    \label{fig:0-state_fidelities}
    \caption{Randomized benchmarking of local Clifford gates on one logical qutrit. A schematic overview of the sequence is presented above the data plot. The protocol consists of an initial fault-tolerant preparation of the $\ket{0}_{\rm L}$-state, followed by the application of \textit{n} transversal logical Clifford gates. The \textit{n}-th gate is chosen to be the inverse of the product of all previous operations. In the absence of errors, the overall Clifford sequence should therefore implement the identity operation. Finally, the system is measured in its $Z$-basis and projected into the code space with a virtual (ideal) round of error correction. 
    Grey points correspond to individual random Clifford sequences. For every randomization, we execute the preparation circuit 100 times. This leads to a stochastic spread in acceptance rate, which results in individual points with limited statistics, causing a large spread in the measured fidelity. 
    The red dashed line shows an exponential fit of the fidelity decaying with the number of gates. While this provides a useful proxy for the average gate performance, the exponential model is not strictly valid in the QEC context, see text for details. The blue dotted line represents a more accurate model based on an individual physical error rate, which does take into account the correction of single errors.}
    \label{fig:1-RB}
\end{figure}

\section{Transversal Clifford gates}
The local Clifford group generators $H_{\rm L}$ and $S_{\rm L}$ are transversal in the Steane code, as shown in Fig.~\ref{fig:0_trap_schematic}b.
They leave the stabilizers, and thus the code space, invariant and transform the logical Pauli operators as $H:X^{\pm 1} \rightarrow Z^{\pm 1}; Z^{\pm 1} \rightarrow X^{\mp 1}$, and $S:X^{\pm 1} \rightarrow (ZX)^{\pm 1}; Z^{\pm 1} \rightarrow Z^{\pm 1}$, where $\pm 1$ denotes the matrix power, such that $X^{1} = X$, $X^{- 1}=X^{\dagger}$, $Z^{1} = Z$, and $Z^{- 1}=Z^{\dagger}$. Each element in the group can be decomposed into its generators and implemented transversally (see Fig.~\ref{fig:0_trap_schematic}b). Consequently, we can perform logical randomized Clifford benchmarking on one logical qutrit, as illustrated in the circuit in Fig.~\ref{fig:1-RB}. To this end, we fault-tolerantly prepare the $\ket{0}_{\rm L}$-state, apply a random sequence of transversal Clifford gates of varying lengths, and append a final inversion gate, so that the overall sequence effectively implements the identity operation.
The ancilla qutrits used for fault-tolerant state preparation are measured only after the Clifford sequence. As the sequence becomes longer, these qutrits are more likely to experience errors while idling, reducing the fraction of accepted runs from about 25\% for 3 Clifford gates to about 8\% for 81 gates. With fewer accepted runs, the statistical uncertainty increases, leading to a larger spread between individual randomizations, as indicated by the gray shaded regions in Fig.~\ref{fig:1-RB}.

A fundamental assumption for extracting an average fidelity estimate from the randomized benchmarking protocol is Markovianity, meaning that the error of each independently chosen gate must depend only on this gate itself, and not any context variable of the protocol run, such as in which position of the circuit it is located. In the context of QEC, the effects of individual gate errors on the logical fidelity cannot generally be considered independently, as the code's ability to correct errors depends on the accumulated error configuration at the time of correction. Consequently, logical dynamics does not necessarily follow a simple exponential decay, and a single average gate fidelity may be insufficient to fully characterize the performance of an error-corrected system~\cite{Combes2017}. This is illustrated in Fig.~\ref{fig:1-RB}, where the exponential fit (red dashed line) does not describe the measured data points within their uncertainties. Nevertheless, the fit yields an average fidelity of F=0.9923(4), which should be interpreted in the context of the specific state-preparation fidelity and range of Clifford sequence lengths considered here.

Alternatively, we model the measured logical fidelity assuming a depolarizing qutrit Pauli noise channel after the state preparation and after each gate in the randomized benchmarking sequence, followed by a final virtual round of (ideal) error correction.
The resulting model (blue dotted line) agrees with the experimental data within their uncertainties. It allows us to estimate the fidelity of a single transversal logical Clifford gate, excluding state-preparation errors and assuming ideal error correction immediately after the gate. This yields a fidelity of  F=0.9990(2) and an effective physical qutrit error rate of  $\text{F$_{\rm \text{phys}}$=0.9970(3)}$. Although the estimated logical gate fidelity exceeds that of the physical gate, this result relies on extrapolating to an idealized regime with perfect error correction that is not realistically accessible in experiments. A more detailed discussion of our model and the interpretation of the results is provided in the Appendix~\ref{sec:rb}.

\begin{figure*}[ht]
    \centering
    \includegraphics[width=0.98\textwidth]{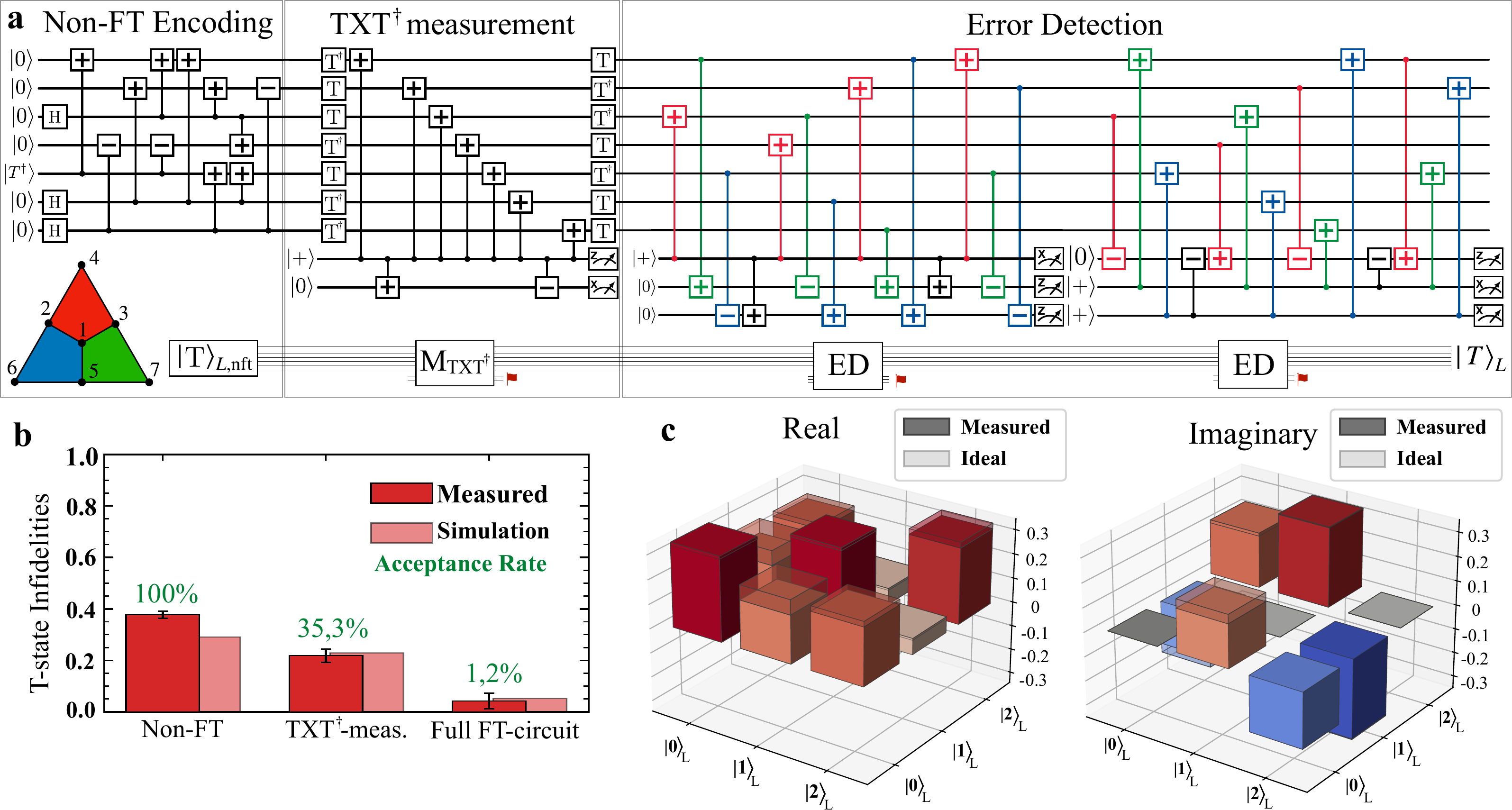}
    \caption{ \textbf{a}: Fault-tolerant $\ket{T}_{\rm L}$ magic-state preparation circuit. A non-FT $\ket{T}_{\rm L}$-state preparation is followed by a logical $TXT^{\dagger}$-measurement using two flag qutrits. In a final round of error detection (ED), all six plaquette stabilizers are measured using $2 \times 3$ ancilla qutrits. \textbf{b}: Logical infidelity with respect to the ideal $\ket{T}_{\rm L}$-state. Each step in the preparation circuit increases the fidelity, up to \SI{0.96(3)}{} for the complete FT protocol. \textbf{c}: Real and imaginary parts of the reconstructed logical density matrix of the $\ket{T}_{\rm L}$-state using logical quantum state tomography and linear inversion reconstruction.}
    \label{fig:4-FT-T-state-preparation}
\end{figure*}

\section{Logical $\ket{T}$-state preparation}
The non-Clifford $T$ gate cannot be transversally implemented in the Steane code, but its FT realization is still needed for a FT universal gate set. One approach to overcoming this limitation is magic-state injection, where non-Clifford operations are realized via injection of fault-tolerantly prepared resource states (see Fig.~\ref{fig:0_trap_schematic}b). These magic states can be obtained through distillation~\cite{Bravyi2005}, which converts a large number of noisy input states into a smaller number of higher-fidelity output states, or through cultivation~\cite{gidney2024magicstatecultivationgrowing, doi:10.1126/science.aee7961}, which improves the fidelity of an encoded magic state using repeated error detection and postselection. Alternatively, they can be prepared directly using fault-tolerant but nondeterministic protocols~\cite{Chamberland_2019, Goto2016, Postler2022}. Another approach is code switching, in which the logical information is transferred to a code that supports the desired non-Clifford gate transversally~\cite{Kubica_2015, Bombin_2015, Pogorelov2025, Bluvstein2026}. Here, we focus on magic-state injection and demonstrate how a magic state can be prepared directly and fault-tolerantly in the seven-qutrit code, thereby providing the key resource needed to realize FT non-Clifford operations and achieve a universal FT gate set.

Following fault-tolerant circuit constructions known for qubits~\cite{Chamberland_2019, Goto2016}, we design a FT magic-state preparation scheme to encode the logical qutrit magic state 
\begin{align}
    \ket{T}_{\rm L} = \frac{1}{\sqrt{3}} \left( \ket{0}_{\rm L} + e^{i\frac{2\pi}{9}} \ket{1}_{\rm L} + e^{- \frac{2\pi}{9}} \ket{2}_{\rm L}\right), \label{eq:logicalTstate}
\end{align} 
defined as the +1 eigenstate of the $T_{\rm L}X_{\rm L}T_{\rm L}^{\dagger}$ operator, in three steps as shown in Fig.~\ref{fig:4-FT-T-state-preparation}. First, the state is encoded non-fault-tolerantly using a single initial physical $\ket{T}$-state. This is followed by a FT measurement of $T_{\rm L}X_{\rm L}T_{\rm L}^{\dagger}$, projecting input states into the operator's eigenbasis. Results corresponding to eigenvalues other than +1 are discarded in order to sort out potential logical errors that a single physical error in the non-FT preparation can potentially produce. An additional flag ensures that this measurement does not propagate any dangerous error from the ancilla qutrits onto the logical state. The final FT error-detection circuit measures the $X$- and $Z$-plaquette stabilizers (Fig.~\ref{fig:0_trap_schematic}a) to identify and reject any states outside the code space, which could have been caused by gate errors in the preceding steps. Only if all of the eight measurements have a trivial outcome, it is guaranteed that the resulting state is the +1 eigenstate of the $T_{\rm L}X_{\rm L}T_{\rm L}^{\dagger}$ gate up to correctable single-qutrit errors, as long as only a single fault has occurred at any position in the circuit.

Experimentally, the price we pay is a reduction of the acceptance rate with each step, dropping from 100\% for the non-FT encoding to 1.2\% after all flag stages. However, the fidelity of the final accepted state with respect to the ideal $\ket{T}_{\rm L}$-state is drastically increased by this, rising from F=0.61(1) in case of the non-FT encoding to F=0.78(3) after the $T_{\rm L}X_{\rm L}T_{\rm L}^{\dagger}$-test and F=0.96(3) for the complete FT circuit, as shown in Fig.~\ref{fig:4-FT-T-state-preparation}b.
This fidelity is measured via logical qutrit state tomography, and the reconstructed logical density matrix of the experimentally prepared magic state is shown in Fig.~\ref{fig:4-FT-T-state-preparation}c, with its entries closely matching the values expected for the ideal magic state.

\section{Discussion and Outlook}
We have demonstrated the fault-tolerant preparation of all $Z$ logical eigenstates and one $X$ eigenstate in the 7-qutrit Steane code, as well as transversal local Clifford gate operations, on a trapped-ion qudit quantum processor.
A key achievement is the fault-tolerant preparation of a logical magic state $\ket{T}_{\rm L}$, which provides a crucial ingredient for logical universal quantum computation. A logical two-qutrit CSUM gate can be implemented transversally using pairwise physical two-qutrit gates, making it comparatively straightforward to realize. Although this operation requires more than the ten qutrits currently available in our quantum processor, with the FT magic state preparation we have demonstrated the more challenging state preparation and non-Clifford single-qutrit gate components needed for universal logical computation. 

These results lay the foundation for fault-tolerant quantum information processing with logical qutrits, extending quantum error correction beyond the protection of qudits encoded as quantum memories in single physical systems. Our work establishes all logical single-qutrit primitives required for fault-tolerant qutrit error correction with high fidelity. This fundamental gateset enables us to run various quantum computing protocols on a logical qutrit level, such as the demonstrated randomized benchmarking or quantum state tomography routines. It can in future work be combined with repeated FT quantum error correction cycles, of which we show here already a single complete round of FT stabilizer measurements used as error detection unit in the FT magic state preparation.

While our demonstration focuses on the seven-qutrit Steane code, its principles and the toolbox developed and shown in this work extend beyond a specific code or local dimension. A broad range of quantum error-correcting codes, including surface codes, color codes, and quantum low-density parity-check codes, admit qudit generalizations~\cite{PhysRevA.87.062338, PhysRevA.92.022312, Spencer_2026}. Moving to higher-dimensional codes offers new opportunities for quantum error correction beyond encoding more information per physical system. In particular, they offer the potential for higher thresholds, lower resource overhead, and enable more efficient FT gate implementations, while retaining access to efficient decoding strategies~\cite{PhysRevA.92.032309, PhysRevX.2.041021, PhysRevLett.113.230501}..

Our demonstration is an essential step towards establishing qudits as a viable alternative to qubits for error-corrected quantum computing. It opens a path towards quantum algorithms that natively decompose into higher-dimensional logical systems~\cite{ Meth2025, Goss2026, john2026, Asaduzzaman2026}, bringing scalable, fault-tolerant qudit quantum computing that genuinely outperforms classical systems closer to realization. 

\section{Methods}

\label{sec:methods}

\subsection{Experimental details}

\subsubsection{The trapped-ion quantum processor}
Measurements were taken on a quantum processor consisting of a chain of ten trapped $^{40}\text{Ca}^+$-ions in a macroscopic linear Paul trap. Information is encoded in the Zeeman sublevels of both the $4\text{S}_{1/2}$ and $3\text{D}_{5/2}$ manifolds, which are connected by a dipole-forbidden, quadrupole-allowed transition at \SI{729}{\nano \meter}. The degeneracy in these manifolds is lifted by an external magnetic field of about \SI{0.638}{\milli \tesla}. This leaves 8 individual spectroscopically addressable levels for information storage, of which the  $3\text{D}_{5/2}(m_j=1/2)$, $4\text{S}_{1/2}(m_j=-1/2)$ and $3\text{D}_{5/2}(m_j=-1/2)$ are used to encode the qutrit states $\ket{0}$, $\ket{1}$ and $\ket{2}$, respectively. Additional levels are used during the optical pumping and sideband cooling stages, as well as for spectroscopically decoupling qutrits during mid-circuit measurements. A detailed description of many of the techniques employed in the experiment can be found in Ref.~\cite{Ringbauer2022}.
\subsubsection{Trapping, Cooling and State Readout}
Using a trap drive frequency of $\Omega = 2 \pi \times$\SI{23.093}{\mega \hertz} and two resonators connecting to the two blade pairs for a dual out-of-phase drive, we work with a radial trap frequency of $\omega = 2 \pi \times$\SI{2.880}{\mega \hertz}.
$^{40}\text{Ca}^+$ has a strong dipole-allowed transition at \SI{397}{\nano \meter} connecting the $4\text{S}_{1/2}$ ground state to the $4\text{P}_{1/2}$ level. With a linewidth of \SI{21.7}{\mega \hertz}, it quickly scatters photons when excited, providing a useful dissipative channel for detection and cooling.
The logical state of an ion can be read out by scattering \SI{397}{\nano \meter} light, which is imaged using an EMCCD camera. An ion scatters light and appears bright only if it is projected onto the ground state, and remains dark if projected onto the $3\text{D}_{5/2}$ manifold. The two states encoded there, $\ket{0}$ and $\ket{2}$, can be distinguished by transferring one of them to the ground state using resonant \SI{729}{\nano \meter} light and repeating the readout process.
We cool the motion of the ions in two stages, using Doppler cooling on the \SI{397}{\nano \meter} transition first, followed by resolved sideband cooling with a \SI{729}{\nano \meter} quadrupole laser. While the Doppler beam cools all modes in all three spatial directions, sideband cooling must be performed for all modes individually, requiring about \SI{20}{\milli \second}. This way, we achieve a mean phonon occupation number of $\Bar{n} = $ \SI{0.02}{\phonons} in the center-of-mass mode used to mediate entangling gates.
 
\subsubsection{Coherent Control}
The levels are optically coupled to each other using a narrow-linewidth \SI{729}{\nano \meter} laser, stabilized to a high-finesse cavity. A laser resonant with an atomic two-level transition results in a rotation on the two-level Bloch sphere that takes the form of $$R(\theta, \phi) = e^{-\frac{\theta}{2}(\text{cos}(\phi)\sigma_x + e^{i \phi}\text{sin}(\phi)\sigma_y)}.$$
A general single-qutrit unitary involving all three levels can be decomposed into Givens rotations, implemented as a series of such two-level couplings using an individually addressed laser beam, and diagonal terms implemented as software phase updates. This decomposition is designed in a way that no direct coupling between the two $3\text{D}_{5/2}$ states is required. 

Shielding our setup with two layers of mu-metal to reduce susceptibility to magnetic fields, we measure a $T_2^* =\SI{216(14)}{\milli \second}$ without dynamical decoupling on the transition used for entangling gates, still limited by slow magnetic field fluctuations. Consequently, transitions with a larger effective Zeeman shift experience shorter coherence times, and those with a small shift longer ones. The states' lifetime in the  $3\text{D}_{5/2}$ manifold is 1.168(9)s.

Entangling operations are implemented through M\o elmer-S\o rensen gates, %~\cite{molmer_sorensen_1999}, 
which can entangle any two ions in a chain using addressed bichromatic laser light, by coupling the ions' electronic state to the chain's motional modes. In the form in which it entangles two states each of ions $i$ and $j$ it can be written as the unitary:
\begin{equation}\label{eq:msgate}
\mathrm{MS}^{ij}(\theta) = e^{-i \frac{\theta}{2} \,\sigma_{x}\otimes\sigma_{x}}
\end{equation}

A maximally entangling gate of two qubits, e.g. a CNOT, can then be implemented with gate angle $\theta = \pi/2$. Typical gate infidelities for isolated maximally entangling gates on two ions are $< 0.01$, slightly varying across ion pairs. This figure increases slightly when entangling all 10 ions to an average of $0.014(4)$ per pair, most likely due to cross-talk effects and phase mismatches across ion pairs. Qudit entangling gates are then composed from multiple applications of such MS-gates~\cite{john2026}.

\subsubsection{Off-Resonant Light Shifts}
Coherent light that couples off-resonantly to an atomic transition with a detuning $\delta$ will cause a frequency shift of $\Delta_{LS} \propto \frac{\Omega^2}{\delta}$. Applying a light pulse of duration $\tau$ induces a relative phase $\phi_{LS} = 2 \pi \cdot \tau  \Delta_{LS}$ between the two energy levels of the transition. Typically in qubit trapped-ion computing, most local operations are resonant with the qubit transition, while entangling MS-gates actively balance the light shift affecting the qubit. However, if additional levels are used to encode a qu\textit{d}it, these spectator levels that do not participate in the M\o lmer-S\o rensen gate will also experience such a shift. These undesired light shifts are sensitive to fluctuations in light intensity and polarization and generally affect each qudit state differently, thus contributing an additional noise channel not present in the qubit setting~\cite{Meth2025}. Nonetheless, the shifts are local and can thus be calibrated and compensated using virtual phase gates.

\subsubsection{Native Qudit Gates and their Decompositions}
Local unitaries on a single qutrit can be decomposed into consecutive two-level operations via Givens rotations. This results in up to three pulses or Bloch-sphere operations for any arbitrary unitary, as well as virtual phases passed onto the three participating levels. A more thorough description of this can be found in the appendix of~\cite{Ringbauer2022}.

All physical entangling operations we use for implementing the qutrit Steane code are CSUM-gates as defined in Eq.~\eqref{eq:CSUM}, or its complex conjugate.
While this gate entangles all three qutrit levels, it can be constructed from MS-gates acting only on the $\ket{1}-\ket{2}$ subspace as well as local operations. This can be done by moving the two-level gate into the $X$-basis, in which it is diagonal, exploiting that the two phase terms imposed by the MS gate are counter-rotating and can thus be driven to be equally spaced from the third state's phase, which remains at 1, thus implementing a qutrit CZ gate. Experimentally, these basis changes can be realized by applying Hadamard gates before and after. A qutrit-Hadamard on the control ion before and after the CZ gate finally turns this into a CSUM. In total, this amounts to
\begin{equation}
\begin{aligned}
    &\mathrm{CSUM}^{ij}_{d=3} =  \\  & {H_{d=3}^j}^\dagger \cdot H_{12}^i\otimes H_{12}^j \cdot \mathrm{MS^{ij}_{12}}\left( \frac{4 \pi}{3}\right) \cdot H_{12}^i\otimes H_{12}^j \cdot H_{d=3}^j ,
\end{aligned}
\end{equation}
where the indices $i$ and $j$ correspond to the ions the gates act on.
Note that this implicitly includes virtual phase operations compensating for the light shift on the third level introduced by the MS-gate.

\subsection{Universal qutrit gate set}
The Pauli operators can be generalized for qudits of dimension $d$. If $d$ is prime, the integers modulo $d$ form a finite field, allowing the stabilizer formalism developed for qubits to be naturally extended to qudits. For qutrits ($d=3$), the Pauli group is generated by
\begin{align}
    X= 
    \begin{pmatrix}
        0&0&1\\
        1&0&0\\
        0&1&0
    \end{pmatrix}, 
    Z= 
    \begin{pmatrix}
        1&0&0\\
        0&\omega&0\\
        0&0&\omega^2
    \end{pmatrix},
\end{align}
where $\omega=\exp(i2\pi / 3)$. They obey the commutation relation $Z X = \omega XZ$. The eigenstates of these operators are given by $\ket{0}$, $\ket{1}$, and $\ket{2}$ for $Z$ and $\ket{+} = (\ket{0} + \ket{1} + \ket{2}) / \sqrt{3}$, $\ket{\omega} = (\ket{0} + \omega \ket{1} + \omega^2 \ket{2}) / \sqrt{3}$, and $\ket{\omega^*} = (\ket{0} + \omega^2 \ket{1} + \omega \ket{2}) / \sqrt{3}$ for $X$.
The Clifford group is generated by 
\begin{align}
    H= \frac{1}{\sqrt{3}}
    \begin{pmatrix}
        1&1&1\\
        1&\omega&\omega^2\\
        1&\omega^2&\omega
    \end{pmatrix}, 
    S= 
    \begin{pmatrix}
        1&0&0\\
        0&\omega&0\\
        0&0&1
    \end{pmatrix}
\end{align}
and the two-qutrit entangling gate
\begin{align}
    \mathrm{CSUM}=\ket{j}\ketbra{j \oplus k}{j}\bra{k}. \label{eq:CSUM}
\end{align}
These gate operations transform Pauli operators to Pauli operators under conjugation. As in the qubit case, the propagation of Pauli errors is thus efficient to track. The Hadamard gate converts $X$- to $Z$-type errors and vice versa,
\begin{align}
    HX &= ZH, \label{eq:hadamard-commutation1}\\
    HZ &= X^{\dagger}H. \label{eq:hadamard-commutation2}
\end{align} 
Similar to the qubit CNOT gate, the CSUM gate copies $X$ errors from control to target and $Z$ errors from target to control qutrit as
\begin{equation}
    \begin{split}
    \mathrm{CSUM}(X^{\alpha}\otimes I) &= (X^{\alpha}\otimes X^{\alpha}) \mathrm{CSUM}, \\
    \mathrm{CSUM} (I \otimes X^{\alpha}) &= (I \otimes X^{\alpha}) \mathrm{CSUM}, \\
    \mathrm{CSUM} (Z^{\alpha} \otimes I) &= (Z^{\alpha} \otimes I) \mathrm{CSUM},\\
    \mathrm{CSUM} (I \otimes Z^{\alpha}) &= (Z^{-\alpha} \otimes Z^{\alpha}) \mathrm{CSUM},
    \end{split} 
\label{eq:csum-propagation}
\end{equation}
where $\alpha \in \{1,2\}$.

Universal control of the qutrit requires a non-Clifford operation such as~\cite{PhysRevA.86.022316, Glaudell_2019}
\begin{align}
    T=\begin{pmatrix}
        1&0&0\\
        0&\omega^{1/3}&0\\
        0&0&\omega^{-1/3}
    \end{pmatrix}, \label{eq:qutrit-Tgate}
\end{align}
which is the qutrit version of the $T$ gate. Under conjugation, it maps Pauli operators to Clifford operators and therefore belongs to the third level of the Clifford hierarchy.

\subsection{The qutrit Steane code}
The qutrit version of the Steane code is a $[[7,1,3]]_d$ code encoding  1 logical qutrit in 7 physical qutrits and has code distance 3. The local dimension of the system is labeled by the index $d=3$. The code is defined by the plaquette stabilizer operators shown in Fig.~\ref{fig:0_trap_schematic}a and logical operators $X_{\rm L}$ and $Z_{\rm L}$ shown in Fig.~\ref{fig:0_trap_schematic}b. As a Calderbank-Shor-Steane (CSS) code~\cite{Steane1996}, it has separate sets of $X$- and $Z$-type stabilizers. The logical qutrit basis states $\ket{i}$, with $i\in {0,1,2}$, are given by
\begin{align}
    \ket{i}_{\rm L} = \frac{1}{\sqrt{|\mathcal{S}_X|}} X_{\rm L}^{i} \sum_{s\in \mathcal{S}_X} s \ket{0000000}, 
\end{align}
where $\mathcal{S}_X = \langle X_1X_2X_3X_4, X_1X_2X_5X_6, X_1X_3X_5X_7 \rangle$ and $|\mathcal{S}_X| = 27$. As in the qubit case logical Clifford gates $H_{\rm L}$, $S_{\rm L}$ and CSUM$_{\rm L}$ are transversal and thus fault tolerant in the code, and generate the Clifford group.

Stabilizer readout in the qutrit Steane code follows the same principles as in the qubit case, allowing single-qutrit errors to be identified and corrected using, e.g., a simple lookup table. 
The stabilizer measurements can be implemented fault-tolerantly in a similar way as in the qubit case. Generalizing the parallel readout scheme proposed in Ref.~\cite{Reichardt_2021} for qubits, we can modify the readout by inserting inverse operators appropriately. This results in the circuit used as the error detection step in Fig.~\ref{fig:4-FT-T-state-preparation}a. In particular, for a proper quantum error correction cycle, the procedure works as follows: If a single error has occurred, either as an incoming error or because a single circuit component has failed, not all ancilla qutrits are measured in their initial state. If such a non-trivial syndrome is measured, the stabilizers catching the respective error type, $X$ or $Z$, are measured one more time without flags. The entire syndrome information of both measurements is then used to correct the error according to a lookup table that can be constructed by propagating all errors on any location in the protocol and assigning them to the respective syndrome outcome.

\subsection{Fault-tolerant universal gate set via magic state preparation}
Magic state distillation and injection with $d$-level quantum systems were already discussed in Refs.~\cite{PhysRevLett.113.230501, PhysRevX.2.041021, PhysRevLett.113.230501, Anwar_2012}. In two-level systems, one distinguishes between two types of resource states: the $\ket{H}$ state is an unequal superposition of the two qubit states and can be used to inject a $\pi/4$ rotation around the $y$-axis, whereas the $\ket{T}$ state is an equal superposition of the qubit states with a relative phase and enables the $T$-gate injection. In general, a qutrit equivalent to the $\ket{H}$ state is distillable~\cite{Anwar_2012}, however, a measurement-based injection of the distilled state would result in a non-unitary gate operation on the system. Instead, magic states need to be converted into {\it phase states}, which are equal superpositions of the three qutrit basis states and only have relative phases with respect to each other, i.e. $\ket{\phi _{\theta, \phi}} = (\ket{0} + e^{i\theta} \ket{1} + e^{i\phi} \ket{2}) / \sqrt{3}$. 

In order to inject the $T^{\dagger}$ gate [Eq.~\eqref{eq:qutrit-Tgate}] one can use the circuit shown in Fig.~\ref{fig:0_trap_schematic}b, where the qutrit analogue of the magic state
\begin{align}
    \ket{T} = \frac{1}{\sqrt{3}} \left( \ket{0} + \omega^{1/3} \ket{1} + \omega^{-1/3} \ket{2}\right), \label{eq:Tstate}
\end{align}
is prepared on an auxiliary qutrit. The data qutrit on which the gate should act is in some arbitrary state $\ket{\psi} = \alpha \ket{0} + \beta \ket{1} + \gamma \ket{2}$. 
The $\mathrm{CSUM}$ gate entangles the two qutrits, and the measurement of the auxiliary qutrit leads to three possible outcomes for the data qutrit state $\ket{\tilde{\psi}}$ listed in the table below, corresponding to $T^{\dagger}\ket{\psi}$, $S^{\dagger} Z T^{\dagger}\ket{\psi}$, and $S T^{\dagger}\ket{\psi}$ for the auxiliary qubit being measured in state $\ket{0}$, $\ket{1}$, and $\ket{2}$, respectively.
\begin{center}
\renewcommand{\arraystretch}{1.5}
    \begin{tabular}{c|c|c}
        auxiliary  & data qubit  & recovery\\
        state & state & operation\\
        \hline
         $\ket{0}$& $\ket{\tilde{\psi}} = \alpha \ket{0} + \omega^{-1/3} \beta \ket{1} + \omega^{1/3} \gamma \ket{2}$ & $I$\\
         
         $\ket{1}$& $\ket{\tilde{\psi}} = \omega^{1/3} \alpha \ket{0} +  \beta \ket{1} + \omega^{-1/3} \gamma \ket{2}$ & $Z^{\dagger}S$\\
         
         $\ket{2}$& $\ket{\tilde{\psi}} = \omega^{-1/3} \alpha \ket{0} + \omega^{1/3} \beta \ket{1} + \gamma \ket{2}$ & $S^{\dagger}$
    \end{tabular}
\end{center}
Depending on the outcome of the measurement, a Clifford recovery operation must be applied to obtain a $T^{\dagger}$ gate. 
Hence, if the logical state $\ket{T}_{\rm L}$ can be prepared fault-tolerantly, both the injection circuit and the required correcting operation ($C$ in Fig.~\ref{fig:0_trap_schematic}b) involve only Clifford gates, which can be implemented transversally in the Steane code. Together, these operations enable a fault-tolerant logical $T^{\dagger}$ gate.

The fault-tolerant logical magic state preparation of $\ket{T}_{\rm L}$ is shown in Fig.~\ref{fig:4-FT-T-state-preparation}a. The $T^{\dagger}$-gate injection circuit then only involves additional transversal ${\rm CSUM}$ gates as well as transversal Clifford operations.  

\bibliography{bibliography.bib}

@article{PhysRevA.87.062338,
  title = {Kitaev's ${\mathbb{Z}}_{d}$-code threshold estimates},
  author = {Duclos-Cianci, Guillaume and Poulin, David},
  journal = {Phys. Rev. A},
  volume = {87},
  issue = {6},
  pages = {062338},
  numpages = {5},
  year = {2013},
  month = {Jun},
  publisher = {American Physical Society},
  doi = {10.1103/PhysRevA.87.062338},
  url = {https://link.aps.org/doi/10.1103/PhysRevA.87.062338}
}

@article{PhysRevA.92.032309,
  title = {Fast fault-tolerant decoder for qubit and qudit surface codes},
  author = {Watson, Fern H. E. and Anwar, Hussain and Browne, Dan E.},
  journal = {Phys. Rev. A},
  volume = {92},
  issue = {3},
  pages = {032309},
  numpages = {12},
  year = {2015},
  month = {Sep},
  publisher = {American Physical Society},
  doi = {10.1103/PhysRevA.92.032309},
  url = {https://link.aps.org/doi/10.1103/PhysRevA.92.032309}
}

@article{PhysRevX.2.041021,
  title = {Magic-State Distillation in All Prime Dimensions Using Quantum Reed-Muller Codes},
  author = {Campbell, Earl T. and Anwar, Hussain and Browne, Dan E.},
  journal = {Phys. Rev. X},
  volume = {2},
  issue = {4},
  pages = {041021},
  numpages = {20},
  year = {2012},
  month = {Dec},
  publisher = {American Physical Society},
  doi = {10.1103/PhysRevX.2.041021},
  url = {https://link.aps.org/doi/10.1103/PhysRevX.2.041021}
}

@article{PhysRevLett.113.230501,
  title = {Enhanced Fault-Tolerant Quantum Computing in $d$-Level Systems},
  author = {Campbell, Earl T.},
  journal = {Phys. Rev. Lett.},
  volume = {113},
  issue = {23},
  pages = {230501},
  numpages = {5},
  year = {2014},
  month = {Dec},
  publisher = {American Physical Society},
  doi = {10.1103/PhysRevLett.113.230501},
  url = {https://link.aps.org/doi/10.1103/PhysRevLett.113.230501}
}

@article{Anwar_2012,
   title={Qutrit magic state distillation},
   volume={14},
   ISSN={1367-2630},
   url={http://dx.doi.org/10.1088/1367-2630/14/6/063006},
   DOI={10.1088/1367-2630/14/6/063006},
   number={6},
   journal={New Journal of Physics},
   publisher={IOP Publishing},
   author={Anwar, Hussain and Campbell, Earl T and Browne, Dan E},
   year={2012},
   month=jun, pages={063006} }

@article{Grassl2003,
author = {Grassl, Markus and Beth, Thomas and R{\"o}tteler, Martin},
title = {On Optimal Quantum Codes},
journal = {International Journal of Quantum Information},
volume = {02},
number = {01},
pages = {55-64},
year = {2004},
doi = {10.1142/S0219749904000079},
URL = {https://doi.org/10.1142/S0219749904000079}
}

@article{PhysRevA.86.022316,
  title = {Qudit versions of the qubit $\ensuremath{\pi}/8$ gate},
  author = {Howard, Mark and Vala, Jiri},
  journal = {Phys. Rev. A},
  volume = {86},
  issue = {2},
  pages = {022316},
  numpages = {10},
  year = {2012},
  month = {Aug},
  publisher = {American Physical Society},
  doi = {10.1103/PhysRevA.86.022316},
  url = {https://link.aps.org/doi/10.1103/PhysRevA.86.022316}
}

@article{Reichardt_2021,
doi = {10.1088/2058-9565/abc6f4},
url = {https://dx.doi.org/10.1088/2058-9565/abc6f4},
year = {2020},
month = {nov},
publisher = {IOP Publishing},
volume = {6},
number = {1},
pages = {015007},
author = {Reichardt, Ben W},
title = {Fault-tolerant quantum error correction for Steane’s seven-qubit color code with few or no extra qubits},
journal = {Quantum Science and Technology}
}

@article{Chamberland_2019,
   title={Fault-tolerant magic state preparation with flag qubits},
   volume={3},
   ISSN={2521-327X},
   url={http://dx.doi.org/10.22331/q-2019-05-20-143},
   DOI={10.22331/q-2019-05-20-143},
   journal={Quantum},
   publisher={Verein zur Forderung des Open Access Publizierens in den Quantenwissenschaften},
   author={Chamberland, Christopher and Cross, Andrew W.},
   year={2019},
   month=may, pages={143} }

@Article{Goto2016,
author={Goto, Hayato},
title={Minimizing resource overheads for fault-tolerant preparation of encoded states of the Steane code},
journal={Scientific Reports},
year={2016},
month={Jan},
day={27},
volume={6},
number={1},
pages={19578},
issn={2045-2322},
doi={10.1038/srep19578},
url={https://doi.org/10.1038/srep19578}
}

@article{PhysRevA.92.022312,
  title = {Qudit color codes and gauge color codes in all spatial dimensions},
  author = {Watson, Fern H. E. and Campbell, Earl T. and Anwar, Hussain and Browne, Dan E.},
  journal = {Phys. Rev. A},
  volume = {92},
  issue = {2},
  pages = {022312},
  numpages = {15},
  year = {2015},
  month = {Aug},
  publisher = {American Physical Society},
  doi = {10.1103/PhysRevA.92.022312},
  url = {https://link.aps.org/doi/10.1103/PhysRevA.92.022312}
}

@article{Glaudell_2019,
   title={Canonical forms for single-qutrit Clifford+<mml:math xmlns:mml=“http://www.w3.org/1998/Math/MathML” display=“inline” overflow=“scroll” id=“d1e19″ altimg=“si5.gif”><mml:mi>T</mml:mi></mml:math> operators},
   volume={406},
   ISSN={0003-4916},
   url={http://dx.doi.org/10.1016/j.aop.2019.04.001},
   DOI={10.1016/j.aop.2019.04.001},
   journal={Annals of Physics},
   publisher={Elsevier BV},
   author={Glaudell, Andrew N. and Ross, Neil J. and Taylor, Jacob M.},
   year={2019},
   month=July, pages={54–70} }

@article{Bravyi2005,
  title = {Universal quantum computation with ideal Clifford gates and noisy ancillas},
  author = {Bravyi, Sergey and Kitaev, Alexei},
  journal = {Phys. Rev. A},
  volume = {71},
  issue = {2},
  pages = {022316},
  numpages = {14},
  year = {2005},
  month = {Feb},
  publisher = {American Physical Society},
  doi = {10.1103/PhysRevA.71.022316},
  url = {https://link.aps.org/doi/10.1103/PhysRevA.71.022316}
}

@article{Kubica_2015,
  title = {Universal transversal gates with color codes: A simplified approach},
  author = {Kubica, Aleksander and Beverland, Michael E.},
  journal = {Phys. Rev. A},
  volume = {91},
  issue = {3},
  pages = {032330},
  numpages = {12},
  year = {2015},
  month = {Mar},
  publisher = {American Physical Society},
  doi = {10.1103/PhysRevA.91.032330},
  url = {https://link.aps.org/doi/10.1103/PhysRevA.91.032330}
}

@article{Bombin_2015,
doi = {10.1088/1367-2630/17/8/083002},
url = {https://doi.org/10.1088/1367-2630/17/8/083002},
year = {2015},
month = {aug},
publisher = {IOP Publishing},
volume = {17},
number = {8},
pages = {083002},
author = {Bombín, Héctor},
title = {Gauge color codes: optimal transversal gates and gauge fixing in topological stabilizer codes},
journal = {New Journal of Physics}
}

@misc{brandl2026,
      title={Quick{Q}udits: A Framework for Efficient Simulation of Noisy Qudit {C}lifford Circuits via an Extended Stabilizer Tableau Formalism}, 
      author={Nina Brandl and Mykyta Cherniak and Johannes Kofler and Richard Kueng},
      year={2026},
      eprint={2603.23641},
      archivePrefix={arXiv},
      primaryClass={quant-ph}
}

@article{Steane1996,
  title = {Error Correcting Codes in Quantum Theory},
  author = {Steane, A. M.},
  journal = {Phys. Rev. Lett.},
  volume = {77},
  issue = {5},
  pages = {793--797},
  numpages = {0},
  year = {1996},
  month = {Jul},
  publisher = {American Physical Society},
  doi = {10.1103/PhysRevLett.77.793},
  url = {https://link.aps.org/doi/10.1103/PhysRevLett.77.793}
}

@article{Gottesman2001,
  title = {Encoding a qubit in an oscillator},
  author = {Gottesman, Daniel and Kitaev, Alexei and Preskill, John},
  journal = {Phys. Rev. A},
  volume = {64},
  issue = {1},
  pages = {012310},
  numpages = {21},
  year = {2001},
  month = {Jun},
  publisher = {American Physical Society},
  doi = {10.1103/PhysRevA.64.012310},
  url = {https://link.aps.org/doi/10.1103/PhysRevA.64.012310}
}

@article{Eastin2009,
  title = {Restrictions on Transversal Encoded Quantum Gate Sets},
  author = {Eastin, Bryan and Knill, Emanuel},
  journal = {Phys. Rev. Lett.},
  volume = {102},
  issue = {11},
  pages = {110502},
  numpages = {4},
  year = {2009},
  month = {Mar},
  publisher = {American Physical Society},
  doi = {10.1103/PhysRevLett.102.110502},
  url = {https://link.aps.org/doi/10.1103/PhysRevLett.102.110502}
}

@misc{Combes2017,
      title={Logical Randomized Benchmarking}, 
      author={Joshua Combes and Christopher Granade and Christopher Ferrie and Steven T. Flammia},
      year={2017},
      eprint={1702.03688},
      archivePrefix={arXiv},
      primaryClass={quant-ph}
}

@article{Brock2025,
author={Brock, Benjamin L. and Singh, Shraddha and Eickbusch, Alec and Sivak, Volodymyr V. and Ding, Andy Z. and Frunzio, Luigi and Girvin, Steven M. and Devoret, Michel H.},
title={Quantum error correction of qudits beyond break-even},
journal={Nature},
year={2025},
month={May},
day={01},
volume={641},
number={8063},
pages={612-618},
issn={1476-4687},
doi={10.1038/s41586-025-08899-y},
url={https://doi.org/10.1038/s41586-025-08899-y}
}

@article{DeBry2026,
author={DeBry, Kyle
and Meister, Nadine
and Valdes Martinez, Agustin
and Bruzewicz, Colin D.
and Shi, Xiaoyang
and Reens, David
and McConnell, Robert
and Chuang, Isaac L.
and Chiaverini, John},
title={Error correction of a logical qubit encoded in a single atomic ion},
journal={Nature Physics},
year={2026},
month={Sep},
day={01},
volume={22},
number={9},
pages={1446-1452},
issn={1745-2481},
doi={10.1038/s41567-026-03315-2},
url={https://doi.org/10.1038/s41567-026-03315-2}
}

@article{john2026,
title={Non-Abelian String-Breaking Dynamics on a Qudit Quantum Computer}, 
author={Manuel John and Keshav Pareek and Peter Tirler and Tim Gollerthan and Michael Meth and Lukas Gerster and Peter Zoller and Daniel González-Cuadra and Torsten V. Zache and Martin Ringbauer},
year={2026},
eprint={2605.05841},
archivePrefix={arXiv},
primaryClass={quant-ph} 
}

@Article{Bluvstein2026,
author={Bluvstein, Dolev
and Geim, Alexandra A.
and Li, Sophie H.
and Evered, Simon J.
and Bonilla Ataides, J. Pablo
and Baranes, Gefen
and Gu, Andi
and Manovitz, Tom
and Xu, Muqing
and Kalinowski, Marcin
and Majidy, Shayan
and Kokail, Christian
and Maskara, Nishad
and Trapp, Elias C.
and Stewart, Luke M.
and Hollerith, Simon
and Zhou, Hengyun
and Gullans, Michael J.
and Yelin, Susanne F.
and Greiner, Markus
and Vuleti{\'{c}}, Vladan
and Cain, Madelyn
and Lukin, Mikhail D.},
title={A fault-tolerant neutral-atom architecture for universal quantum computation},
journal={Nature},
year={2026},
month={Jan},
day={01},
volume={649},
number={8095},
pages={39-46},
issn={1476-4687},
doi={10.1038/s41586-025-09848-5},
url={https://doi.org/10.1038/s41586-025-09848-5}
}

@Article{Acharya2025,
author={{Google Quantum AI and Collaborators}},
title={Quantum error correction below the surface code threshold},
journal={Nature},
year={2025},
month={Feb},
day={01},
volume={638},
number={8052},
pages={920-926},
issn={1476-4687},
doi={10.1038/s41586-024-08449-y},
url={https://doi.org/10.1038/s41586-024-08449-y}
}

@Article{Krinner2022,
author={Krinner, Sebastian
and Lacroix, Nathan
and Remm, Ants
and Di Paolo, Agustin
and Genois, Elie
and Leroux, Catherine
and Hellings, Christoph
and Lazar, Stefania
and Swiadek, Francois
and Herrmann, Johannes
and Norris, Graham J.
and Andersen, Christian Kraglund
and M{\"u}ller, Markus
and Blais, Alexandre
and Eichler, Christopher
and Wallraff, Andreas},
title={Realizing repeated quantum error correction in a distance-three surface code},
journal={Nature},
year={2022},
month={May},
day={01},
volume={605},
number={7911},
pages={669-674},
issn={1476-4687},
doi={10.1038/s41586-022-04566-8},
url={https://doi.org/10.1038/s41586-022-04566-8}
}

@Article{Paetznick2026,
author={Paetznick, A. and Reichardt, B. W. and Silva, M. P. da and Ryan-Anderson, C. and Aasen, D. and Bello-Rivas, J. M. and Campora, J. P. and Chao, R. and Chernoguzov, A. and van Dam, W. and Dreiling, J. M. and Foltz, C. and Frachon, F. and Gaebler, J. P. and Gatterman, T. M. and Grans-Samuelsson, L. and Gresh, D. and Hayes, D. and Hewitt, N. and Holliman, C. and Horst, C. V. and Johansen, J. and Lucchetti, D. and Matsuoka, Y. and Mills, M. and Moses, S. A. and Neyenhuis, B. and Paz, A. and Pino, J. and Siegfried, P. and Sundaram, A. and Tom, D. and Wang, Z. and Wernli, S. J. and Zanner, M. and Stutz, R. P. and Svore, K. M.},
title={Improved quantum processor logical error rates via correction and detection},
journal={Nature},
year={2026},
month={Jun},
day={01},
volume={654},
number={8118},
pages={349-355},
issn={1476-4687},
doi={10.1038/s41586-026-10628-y},
url={https://doi.org/10.1038/s41586-026-10628-y}
}

@article{Abobeih2022,
   author = {M. H. Abobeih and Y. Wang and J. Randall and S. J.H. Loenen and C. E. Bradley and M. Markham and D. J. Twitchen and B. M. Terhal and T. H. Taminiau},
   doi = {10.1038/s41586-022-04819-6},
   issn = {14764687},
   issue = {7916},
   journal = {Nature},
   month = {6},
   pages = {884-889},
   pmid = {35512730},
   publisher = {Nature Research},
   title = {Fault-tolerant operation of a logical qubit in a diamond quantum processor},
   volume = {606},
   year = {2022}
}

@article{Egan2021,
   author = {Laird Egan and Dripto M. Debroy and Crystal Noel and Andrew Risinger and Daiwei Zhu and Debopriyo Biswas and Michael Newman and Muyuan Li and Kenneth R. Brown and Marko Cetina and Christopher Monroe},
   doi = {10.1038/s41586-021-03928-y},
   issn = {14764687},
   issue = {7880},
   journal = {Nature},
   month = {10},
   pages = {281-286},
   pmid = {34608286},
   publisher = {Nature Research},
   title = {Fault-tolerant control of an error-corrected qubit},
   volume = {598},
   year = {2021}
}

@article{Erhard2021,
   author = {Alexander Erhard and Hendrik Poulsen Nautrup and Michael Meth and Lukas Postler and Roman Stricker and Martin Stadler and Vlad Negnevitsky and Martin Ringbauer and Philipp Schindler and Hans J. Briegel and Rainer Blatt and Nicolai Friis and Thomas Monz},
   doi = {10.1038/s41586-020-03079-6},
   issn = {14764687},
   issue = {7841},
   journal = {Nature},
   month = {1},
   pages = {220-224},
   pmid = {33442044},
   publisher = {Nature Research},
   title = {Entangling logical qubits with lattice surgery},
   volume = {589},
   year = {2021}
}

@article{Postler2022,
   author = {Lukas Postler and Sascha Heußen and Ivan Pogorelov and Manuel Rispler and Thomas Feldker and Michael Meth and Christian D. Marciniak and Roman Stricker and Martin Ringbauer and Rainer Blatt and Philipp Schindler and Markus Müller and Thomas Monz},
   doi = {10.1038/s41586-022-04721-1},
   issn = {14764687},
   issue = {7911},
   journal = {Nature},
   month = {5},
   pages = {675-680},
   pmid = {35614250},
   publisher = {Nature Research},
   title = {Demonstration of fault-tolerant universal quantum gate operations},
   volume = {605},
   year = {2022}
}

@article{Chi2022,
    author={Chi, Yulin
    and Huang, Jieshan
    and Zhang, Zhanchuan
    and Mao, Jun
    and Zhou, Zinan
    and Chen, Xiaojiong
    and Zhai, Chonghao
    and Bao, Jueming
    and Dai, Tianxiang
    and Yuan, Huihong
    and Zhang, Ming
    and Dai, Daoxin
    and Tang, Bo
    and Yang, Yan
    and Li, Zhihua
    and Ding, Yunhong
    and Oxenl{\o}we, Leif K.
    and Thompson, Mark G.
    and O'Brien, Jeremy L.
    and Li, Yan
    and Gong, Qihuang
    and Wang, Jianwei},
    title={A programmable qudit-based quantum processor},
    journal={Nature Communications},
    year={2022},
    month={Mar},
    day={04},
    volume={13},
    number={1},
    pages={1166},
    issn={2041-1723},
    doi={10.1038/s41467-022-28767-x},
    url={https://doi.org/10.1038/s41467-022-28767-x}
}

@article{Ringbauer2022,
   author = {Martin Ringbauer and Michael Meth and Lukas Postler and Roman Stricker and Rainer Blatt and Philipp Schindler and Thomas Monz},
   doi = {10.1038/s41567-022-01658-0},
   issn = {17452481},
   issue = {9},
   journal = {Nature Physics},
   month = {9},
   pages = {1053-1057},
   publisher = {Nature Research},
   title = {A universal qudit quantum processor with trapped ions},
   volume = {18},
   year = {2022},
}

@article{Yu2025,
    author={Yu, Xi
    and Wilhelm, Benjamin
    and Holmes, Danielle
    and Vaartjes, Arjen
    and Schwienbacher, Daniel
    and Nurizzo, Martin
    and Kringh{\o}j, Anders
    and Blankenstein, Mark R. van
    and Jakob, Alexander M.
    and Gupta, Pragati
    and Hudson, Fay E.
    and Itoh, Kohei M.
    and Murray, Riley J.
    and Blume-Kohout, Robin
    and Ladd, Thaddeus D.
    and Anand, Namit
    and Dzurak, Andrew S.
    and Sanders, Barry C.
    and Jamieson, David N.
    and Morello, Andrea},
    title = {Schrödinger cat states of a nuclear spin qudit in silicon},journal={Nature Physics},
    year={2025},
    month={Mar},
    day={01},
    volume={21},
    number={3},
    pages={362-367},
    issn={1745-2481},
    doi={10.1038/s41567-024-02745-0},
    url={https://doi.org/10.1038/s41567-024-02745-0}
}

@article{Morvan2021,
  title = {Qutrit Randomized Benchmarking},
  author = {Morvan, A. and Ramasesh, V. V. and Blok, M. S. and Kreikebaum, J. M. and O'Brien, K. and Chen, L. and Mitchell, B. K. and Naik, R. K. and Santiago, D. I. and Siddiqi, I.},
  journal = {Phys. Rev. Lett.},
  volume = {126},
  issue = {21},
  pages = {210504},
  numpages = {6},
  year = {2021},
  month = {May},
  publisher = {American Physical Society},
  doi = {10.1103/PhysRevLett.126.210504},
  url = {https://link.aps.org/doi/10.1103/PhysRevLett.126.210504}
}

@misc{lavasani2026,
      author={Ali Lavasani and Adam P. Reed and Joan M. Dreiling and Annie J. Park and Michael Mills and Matteo Puviani and Gabriel Matos and M. S. Allman and Noah Berthusen and Natalie C. Brown and Victor E. Colussi and Brian Estey and Cameron Foltz and Daniel Gresh and Reza Haghshenas and Akhil Isanaka and Nikhil Kotibhaskar and Anthony Ransford and Ciarán Ryan-Anderson and Brian Neyenhuis and David Hayes and Tzvetan S. Metodi and Andrew C. Potter},
      title={Quantum error correction well below the color code threshold}, 
      year={2026},
      eprint={2610.24077},
      archivePrefix={arXiv},
      primaryClass={quant-ph}
}

@article{
wang2018,
author = {Jianwei Wang  and Stefano Paesani  and Yunhong Ding  and Raffaele Santagati  and Paul Skrzypczyk  and Alexia Salavrakos  and Jordi Tura  and Remigiusz Augusiak  and Laura Mančinska  and Davide Bacco  and Damien Bonneau  and Joshua W. Silverstone  and Qihuang Gong  and Antonio Acín  and Karsten Rottwitt  and Leif K. Oxenløwe  and Jeremy L. O’Brien  and Anthony Laing  and Mark G. Thompson },
title = {Multidimensional quantum entanglement with large-scale integrated optics},
journal = {Science},
volume = {360},
number = {6386},
pages = {285-291},
year = {2018},
doi = {10.1126/science.aar7053},
URL = {https://www.science.org/doi/abs/10.1126/science.aar7053}
}

@article{Low2026,
author={Low, Pei Jiang
and Zutt, Nicholas C. F.
and Tathed, Gaurav A.
and Senko, Crystal},
title={Quantum logic operations and algorithms in a single 25-level atomic qudit},
journal={Nature Communications},
year={2026},
month={Jun},
day={02},
volume={17},
number={1},
pages={7098},
issn={2041-1723},
doi={10.1038/s41467-026-72662-8},
url={https://doi.org/10.1038/s41467-026-72662-8}
}

@article{Rojkov2024,
  title = {Two-Qubit Operations for Finite-Energy Gottesman-Kitaev-Preskill Encodings},
  author = {Rojkov, Ivan and R{\"o}ggla, Paul Moser and Wagener, Martin and Fontbot{\'e}-Schmidt, Moritz and Welte, Stephan and Home, Jonathan and Reiter, Florentin},
  journal = {Phys. Rev. Lett.},
  volume = {133},
  issue = {10},
  pages = {100601},
  numpages = {7},
  year = {2024},
  month = {Sep},
  publisher = {American Physical Society},
  doi = {10.1103/PhysRevLett.133.100601},
  url = {https://link.aps.org/doi/10.1103/PhysRevLett.133.100601}
}

@article{Spencer_2026,
   title={Qudit low-density parity-check codes},
   volume={10},
   ISSN={2521-327X},
   url={http://dx.doi.org/10.22331/q-2026-03-13-2023},
   DOI={10.22331/q-2026-03-13-2023},
   journal={Quantum},
   publisher={Verein zur Forderung des Open Access Publizierens in den Quantenwissenschaften},
   author={Spencer, Daniel J. and Tanggara, Andrew and Haug, Tobias and Khu, Derek and Bharti, Kishor},
   year={2026},
   month=Mar, pages={2023} }

@article{Meth2025,
author={Meth, Michael and Zhang, Jinglei and Haase, Jan F. and Edmunds, Claire and Postler, Lukas and Jena, Andrew J. and Steiner, Alex and Dellantonio, Luca and Blatt, Rainer and Zoller, Peter and Monz, Thomas and Schindler, Philipp and Muschik, Christine and Ringbauer, Martin},
title={Simulating two-dimensional lattice gauge theories on a qudit quantum computer},
journal={Nature Physics},
year={2025},
month={Apr},
day={01},
volume={21},
number={4},
pages={570-576},
issn={1745-2481},
doi={10.1038/s41567-025-02797-w},
url={https://doi.org/10.1038/s41567-025-02797-w}
}

@misc{Goss2026,
      title={A Qutrit Time Crystal Stabilized with Native Chiral Interactions}, 
      author={Noah Goss and Nishchay Suri and Brian Marinelli and Larry Chen and Akel Hashim and Sajant Anand and Alexis Morvan and Ravi K. Naik and Ermal Rrapaj and David I. Santiago and Wibe de Jong and Norman Y. Yao and Joel E. Moore and Irfan Siddiqi},
      year={2026},
      eprint={2605.14293},
      archivePrefix={arXiv},
      primaryClass={quant-ph} 
}

@article{Asaduzzaman2026,
  title = {Hybrid analog-digital simulation of the Abelian Higgs model with transmon qutrits},
  author = {Asaduzzaman, Muhammad and Parker, Rayleigh W. and Goss, Noah and Mohamed, Ahmed I. and Neiderbach, Max and Ozzello, Zane and Naik, Ravi K. and Kemper, Alexander F. and Siddiqi, Irfan and Meurice, Yannick and Blok, Machiel S.},
  journal = {Phys. Rev. A},
  volume = {114},
  issue = {4},
  pages = {042462},
  numpages = {29},
  year = {2026},
  month = {Oct},
  publisher = {American Physical Society},
  doi = {10.1103/r632-zl7s},
  url = {https://link.aps.org/doi/10.1103/r632-zl7s}
}

@article{Wang2020,
AUTHOR={Wang, Yuchen  and Hu, Zixuan  and Sanders, Barry C.  and Kais, Sabre },
TITLE={Qudits and High-Dimensional Quantum Computing},
JOURNAL={Frontiers in Physics},
VOLUME={8},
pages={589504},
YEAR={2020},
DOI={10.3389/fphy.2020.589504},
URL={https://www.frontiersin.org/journals/physics/articles/10.3389/fphy.2020.589504},
ISSN={2296-424X}
}

@misc{tham2026breakevendemonstrationquantumlowdensity,
      title={Breakeven demonstration of quantum low-density parity-check codes}, 
      author={Edwin Tham and Michael L. Goldman and Shantanu Debnath and Ashay N. Patel and Jyothi Saraladevi and Jason Nguyen and Erik Nielsen and Neal Pisenti and Kenneth Wright and John Gamble and Nicolas Delfosse},
      year={2026},
      eprint={2606.06455},
      archivePrefix={arXiv},
      primaryClass={quant-ph}
}

@article{PhysRevLett.121.050502,
  title = {Quantum Error Correction with Only Two Extra Qubits},
  author = {Chao, Rui and Reichardt, Ben W.},
  journal = {Phys. Rev. Lett.},
  volume = {121},
  issue = {5},
  pages = {050502},
  numpages = {5},
  year = {2018},
  month = {Aug},
  publisher = {American Physical Society},
  doi = {10.1103/PhysRevLett.121.050502},
  url = {https://link.aps.org/doi/10.1103/PhysRevLett.121.050502}
}

@misc{gidney2024magicstatecultivationgrowing,
      title={Magic state cultivation: growing T states as cheap as CNOT gates}, 
      author={Craig Gidney and Noah Shutty and Cody Jones},
      year={2024},
      eprint={2409.17595},
      archivePrefix={arXiv},
      primaryClass={quant-ph}
}

@article{
doi:10.1126/science.aee7961,
author = {E. Rosenfeld  and others},
title = {Magic state cultivation on a superconducting quantum processor},
journal = {Science},
volume = {0},
number = {0},
pages = {eaee7961},
year = {2026},
doi = {10.1126/science.aee7961},
URL = {https://www.science.org/doi/abs/10.1126/science.aee7961}
}

@Article{Pogorelov2025,
author={Pogorelov, Ivan
and Butt, Friederike
and Postler, Lukas
and Marciniak, Christian D.
and Schindler, Philipp
and M{\"u}ller, Markus
and Monz, Thomas},
title={Experimental fault-tolerant code switching},
journal={Nature Physics},
year={2025},
month={Feb},
day={01},
volume={21},
number={2},
pages={298-303},
issn={1745-2481},
doi={10.1038/s41567-024-02727-2},
url={https://doi.org/10.1038/s41567-024-02727-2}
}

\appendix
\newpage
\begin{center}
  {\large\bfseries APPENDIX}\\[0.5em]
\end{center}
%\vspace{0.em}

\section{Error Model and Simulations}
\label{sec:error-model}

In this work, we show two different types of simulation of the experiments we run. For Clifford-only circuits we use a stabilizer-tableau formalism based on the \textit{quickqudits}-package presented in \cite{brandl2026}. In contrast, we perform python-based state-vector simulations of the $T$-state preparation circuit as it contains non-Clifford operations.

For the error model, we assume that the entangling CSUM gates are the most noisy error locations. They suffer from local noise on each of the participating qutrits, due to imperfections in the underlying two-level MS gate as well as drifts and calibration mismatches in the off-resonant light shift introduced on the third qutrit level. A detailed characterization of the structure of experimental noise in qutrits can be complicated, and as of now there has not been such a study of noise models on the experimental setup used. While in principle the various physical noise mechanisms could be modeled in more detail, e.g. light-shifts causing $Z$-type-errors on the control qudit and $X$-type-errors on the target qutrit, we simply assume an effective local depolarizing Pauli-noise channel of the form:

\begin{equation}
    \mathcal{E}(\rho) = (1-p) \rho + \frac{p}{K} \sum_{k=1}^{K} P_k \rho P_k^{\dagger} . \label{eq:error-on-single-qubit-gates}
\end{equation}
where $p$ corresponds to the probability that an error has occurred, and $P_k$ with $k=1,\dots, K$ are the qutrit Pauli operators. In the qutrit case all non-trivial combinations of $X$ and $Z$ lead to $K=8$ possible Pauli errors and each occurs with probability $p/8$. Describing noise in the Pauli basis is convenient as it makes it easy to track errors propagating through the circuits, and we observe good agreement with our experimental data.

We use this depolarizing noise channel to model errors during state preparation and readout at the beginning and end of the circuit, assigning an error likelihood of $p = 0.01$ each. Errors on the local operations at the beginning and end of the circuit are included in this value.
For two-qutrit gates, we consider both local uncorrelated depolarizing errors acting on both qutrits in the CSUM gate independently, as well as correlated two-qutrit Pauli errors acting on both simultaneously. We fit the CSUM gate error to the experimental data of the state preparation circuit, resulting in a depolarizing error rate of 0.031(1) for the independent local noise and 0.0155(3) for the correlated two-qutrit Pauli noise. We observe that a correlated error of roughly half of the local error fits the data best, so in order to reduce the number of free parameters their relation is forced to $p_{\text{local}} = 2 p_{\text{correlated}}$. These values are then used for the simulations shown of all other experiments too, such as magic state injection.

\section{Decoder}
\label{sec:decoder}

We perform decoding based on the measured syndromes. Syndromes associated with weight-one errors relative to any state in the logical subspace can be corrected by applying the corresponding correction operator. In the qutrit case, however, certain syndromes arise from genuine weight-two errors with respect to the entire code space. One example of such an error for the encoded zero state is shown in Fig.~2a of the main text. 
Since different weight-two errors can produce the same syndrome, the corresponding correction is not necessarily unique. For each syndrome that cannot be attributed to a weight-one error, we identify the set of weight-two errors consistent with the measurement outcome. 

\begin{figure*}[ht]
    \centering
    \includegraphics[width=0.95\textwidth]{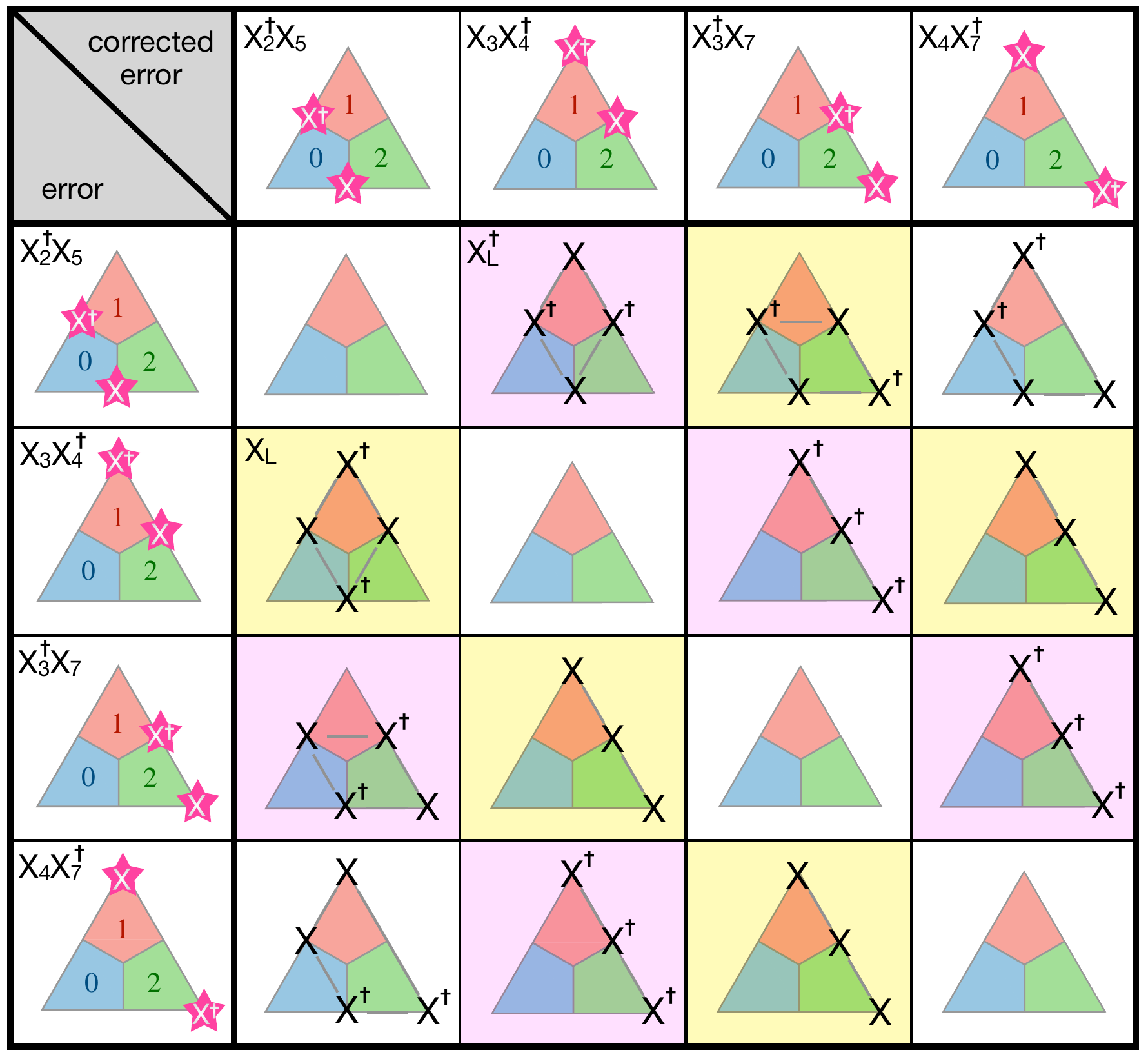}
    \caption{Weight-two $X$-faults that cause the ancillas of the red, green and blue $Z$-stabilizer measurement to be in states $\ket{1}$, $\ket{2}$, and $\ket{0}$ respectively, and their closest non-trivial stabilizer and logicals.}
    \label{fig:error-classification}
\end{figure*}

Unlike in the qubit case there are now three possible logical states into which a given syndrome can be classified. For example, the measurement of the ancilla in state $\ket{1}$ for the red, $\ket{2}$ for the green, and $\ket{0}$ for the blue $Z$-stabilizers can be induced by errors $X_2^{\dagger} X_5$, $X_3 X_4^{\dagger}$, $X_3^{\dagger} X_7$, or $X_4 X_7^{\dagger}$ which are shown in Fig.~\ref{fig:error-classification}. For each possible error corresponding to the syndrome, we can apply each possible correction and find the resulting operation. In Fig.~\ref{fig:error-classification} the white cells mark stabilizer elements (meaning the right correction is applied), the yellow cells mark $X_L$-operators and the pink cells mark $X_L^{\dagger}$-operators. 

For example, let us assume that the error $X_2^{\dagger} X_5$ has occurred (first error in the table). Since we can only rely on the syndrome information (1,2,0), we do not know by which correction we would recover the state. We can correct the error with a correction corresponding to all possible error candidates (columns). If we correct the error with its inverse, the recovery is successful and the error is removed. However, if we correct with $X_3^{\dagger} X_4$, which corresponds to the inverse of the second possible error, this ultimately results in $X_2^{\dagger} X_3^{\dagger} X_4 X_5$ which is a logical $X^{\dagger}$-error (pink cell in the table). If we correct instead with $X_3 X_7^{\dagger}$ this leads to the logical $X$-error $X_2^{\dagger} X_3 X_5 X_7^{\dagger}$ (yellow cell). On the other hand, correcting with $X_4^{\dagger} X_7$ leads to the stabilizer $X_2^{\dagger} X_4^{\dagger} X_5 X_7$. In summary, two of the four possible corrections lead to the right recovery operation up to a stabilizer. We can group these two errors into the same error class, as they are equivalent up to a stabilizer. We combine each error with all recovery operations, and summarize the results in Fig.~\ref{fig:error-classification}. If we act on the encoded state with a depolarizing noise channel, Eq.~\eqref{eq:error-on-single-qubit-gates}, we overall find that given the syndrome (1,2,0), there is a 50\% chance that the error was in $\{X_2^{\dagger} X_5, X_4 X_7^{\dagger}\}$, which corresponds to the same error up to a stabilizer; there is a 25\% chance that the error was $X_3 X_4^{\dagger}$ and a 25\% that the error was $X_3^{\dagger} X_7$. If we thus apply the recovery operation $X_2 X_5^{\dagger}$, we recover to the right state with 50\% probability, we make a logical error $X^{\dagger}_L$ or $X_L$ with 25\% probability each. 

The underlying assumption of each of the four errors occurring with the same or at least similar likelihood is useful in the case of unknown circuits, or circuits with little structure, but in general not necessarily true. There might be circuits where the specific error likelihoods vary, and considering them would lead to a decoding performance superior to counting each possible error with the same weight. For small circuits such as the state preparation circuits we consider, this can be done using a simulation of the circuit with a physical noise model. 
We consider three decoding strategies for selecting a correction:

The \textit{50-50 decoder}: For each measured syndrome, we identify the compatible weight-two errors, assigning equal probability to each. We then group these errors into equivalence classes under the stabilizer group and select the class with the highest total probability to determine the correction. This is most useful when no particular structure is assumed in the circuit, and all errors are considered roughly equally likely.

The \textit{bare circuit-informed decoder}: This decoder uses error statistics obtained from simulations of the encoding circuit using circuit-level noise. For each syndrome, we count how often each corresponding error occurs and select the most frequently observed error to determine the correction.

The \textit{classifying circuit-informed decoder}: This decoder extends the bare circuit-informed decoder by grouping errors that are equivalent up to stabilizer operators into a single error class. The occurrence counts of all errors within each class are summed, and the class with the highest total count is selected to determine the correction.

\section{Code capacity of the qutrit Steane code}
\label{sec:code-capacity}

In the following, we investigate the logical error rate in the Steane code for the qubit as well as for the qutrit case. For this, we assume depolarizing Pauli noise, as in Eq.~\eqref{eq:error-on-single-qubit-gates} on each qubit (qutrit) individually, where $P_k$, $k=1,\dots, K$ are the qubit (qutrit) Pauli operators, respectively, and the physical error rate $p$ corresponds to the probability that an error has occurred. Note that in the qubit case $K=3$, i.e., there exist only three different Pauli operators $X$, $Y$, and $Z$, each occurring with probability $p/3$. In the qutrit case, all non-trivial combinations of $X$ and $Z$ lead to $K=8$ possible Pauli errors, and each occurs with probability $p/8$. 

For small physical error probabilities, the logical error rate is dominated by weight-two faults. For qubits, there are 189 weight-two errors, of which 42 are correctable (one $X$ and one $Z$ error). This leaves 147 non-correctable weight-two faults. The logical error can therefore be estimated by $p_L \approx 147 (p/3)^2 + \Order[]{p^3} \approx 16.3 p^2$. For qutrits, there are 1344 weight-two faults, of which 168 contain one $X$-type and one $Z$-type fault that are correctable. This yields an upper bound for the error probability of $p_L \approx 1176 (p/8)^2 + \Order[]{p^3} \approx 18.4 p^2$. As discussed in the Decoder Section, there exist error syndromes triggered by weight-two faults which do not belong to weight-one faults. Thus, it is possible to correct for more weight-two errors as compared to the qubit case. 

\begin{figure}[h]
    \centering
    \includegraphics[width=0.4\textwidth]{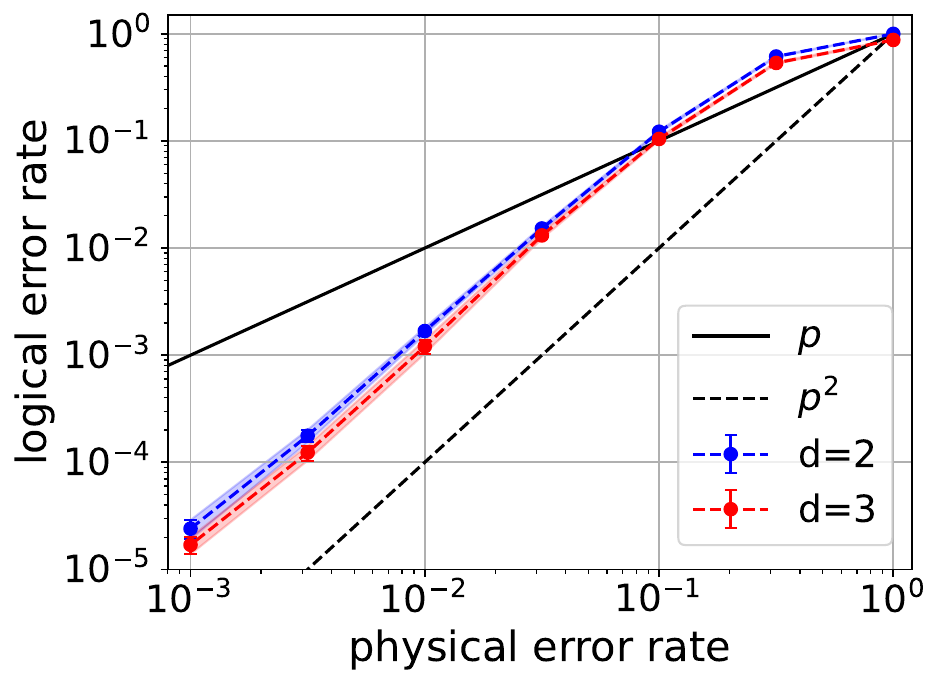}
    \caption{Simulation of the code capacity of the qubit (blue) and qutrit (red) Steane code. The solid and dashed black lines indicate $p$ and $p^2$ scaling.}
    \label{fig:code_capacity_compare}
\end{figure}

We simulate the code capacity of the Steane code for the qubit and qutrit case using Monte-Carlo sampling and show the logical error rate depending on the physical error rate as blue and red data points in Fig.~\ref{fig:code_capacity_compare}. Subsequently, we apply a lookup table decoder in the qubit case, accounting for any single-qubit fault. In the qutrit case, we decode any syndrome that belongs to a single-weight fault using a look-up table. For syndromes that do not correspond to a single-qutrit fault, we select the most probable stabilizer-equivalent weight-two error, as described in the decoder section labeled by the 50-50-decoder. 

Although qubit and qutrit Pauli error rates are not directly comparable, we present both results in the same plot in Fig.~\ref{fig:code_capacity_compare}. In both cases, we find a quadratic scaling in the physical error rate as expected. We find that the curve is slightly shifted for qutrits, leading to a slightly higher pseudo-threshold. The logical error estimate of $p_L = 16.3 p^2$ agrees well with the qubit data points (not shown). The logical qutrit error rate performs better than the upper bound calculated above. Choosing the 50-50-decoder, we can fit $p_L = 12.5p^2$, not shown in the figure. Our results indicate that error thresholds for qutrits remain comparable to the qubit case while more information is encoded in the system. In contrast to the qubit case, the larger number of possible syndromes allows additional errors to be identified and corrected using an appropriate decoder.

\section{Modeling Logical Randomized Benchmarking}
\label{sec:rb}

Typically, experimental noise can have some structure in which not all error channels occur equally likely, and consequently a quantum gate's observed performance or fidelity will differ depending on which input state it is applied to. One possible way to characterize gates is to estimate their effective quantum channel through schemes like process tomography, which provides complete information but is experimentally costly and sensitive to SPAM errors. An alternative approach (for some gate sets, such as Clifford gates) is randomized benchmarking. Here, long sequences of random Clifford gates are used to ``twirl" all noise channels into depolarizing noise, which allows to give one number quantifying the average gate fidelity in the set. While this loses information about the noise structure, it is experimentally accessible and returns a clear figure of merit. 

The underlying assumption is that errors occur independently and with equal probability at each gate. However, quantum error correction changes the resulting error distribution, as some errors are corrected or detected and discarded, while others remain undetected. Consequently, errors that affect the final logical state are no longer equally likely. 

Any measurement of the logical qutrit actually returns a distribution of states either in the computational subspace or outside of it, much like Fig.~2b in the main text. In order to assign it to a clear logical measurement result within the logical space, a virtual round of error correction is performed. This can correct for single physical errors, but as errors accumulate and two or more physical errors occur, they can potentially result in a logical one. Consequently, when looking at randomized benchmarking at a logical level, the likelihood of logical errors cannot be treated as Markovian anymore. This is in conflict with a core assumption of the randomized benchmarking protocol and thus limits our ability to extract a single average gate fidelity estimate. This effect is known and discussed in literature, while methods to benchmark logical gates remain an active area of research~\cite{Combes2017, Paetznick2026}.
In our measured data, this manifests as a deviation from an ideal exponential decay, exhibiting a feature on top that might look gaussian---initial gates add less infidelity, followed by a region of faster decay per gate.

In our experiment, the randomized benchmarking sequence follows the FT preparation of the logical $\ket{0}$-state, without additional rounds of error correction between the two or between individual logical Clifford gates. Although the observed decay is well approximated by an exponential function, this behavior is not generally guaranteed when using QEC. The extracted fidelity should therefore be interpreted as an effective measure of gate performance under the specific experimental conditions of our benchmarking protocol.

Alternatively, one can compute how local, physical depolarizing noise accumulates during the randomized benchmarking protocol until the virtual round of ideal error correction is performed. Note that the random Clifford operations on a physical level still ensure that the assumption of physical depolarizing noise is correct.
To estimate the logical error probability, we count all possible error combinations that result in a single-qutrit error at the end of the circuit. After FT initialization of the zero-state, we effectively model the probability of an error in a single qutrit as $p$. For the seven qutrits in the code, the probability that no error has occurred is then $(1-p)^7$, and the probability that a single error has occurred is $7p(1-p)^6$. Assuming a perfect decoding step that corrects for all single-qutrit errors, the logical error probability is approximated by $p_L = 1-(1-p)^7 - 7p(1-p)^6$. After one single-qutrit gate, the probability of a single resulting error consists of two scenarios: i) Only one error has occurred, either in the initialization or in the gate, each with probability $(1-p)p$. ii) An error has occurred in both steps, but the first error is not cancelled by an error in the second step. Since there are seven out of eight such qutrit Pauli errors, the probability is $7p^2/8$. Overall, this leads to the probability for a resulting single-qutrit error of $r_1=2(1-p)p + 7p^2/8$ after one gate. The logical error can then be approximated by $p_{\rm L} = 1-(1-r_1)^7 - 7 r_1(1-r_1)^6$. Continuing this counting of non-trivial errors for $m$ gates, we derive the logical error probability $p_{\rm L}$ after a final round of error correction, assuming that all single-qutrit errors are corrected as
\begin{align}
    p_{\rm L} = 1 - (1-r_m)^7 -7r(1-r_m)^6, \label{eq:fit-error-RBexp}
\end{align}
where $$r_m = \frac{8}{9} \left(1-(1-\frac{9p}{8})^{m+1} \right).$$
Here $p$ is the error probability of a physical Clifford gate under depolarizing noise, and $m$ the number of applied logical Clifford gates. This provides a model for fitting the logical randomized benchmarking data in the presence of error correction. The model accounts for faulty state preparation by considering an unknown amount of physical Clifford gates carried out before the start of the benchmarking sequence as a free parameter. This way, the state preparation error can be considered as it builds up across correctable and uncorrectable errors. 

Using Eq.~\eqref{eq:fit-error-RBexp}, we fit an average \textit{effective physical} fidelity of $\textbf{0.9970(3)}$ to the data of Fig.~3. 
In contrast to randomized benchmarking on a physical level, for the logical qutrit, we do not extract an effective logical depolarizing channel but fit the measured logical error probabilities to an effective physical depolarizing error rate. Although the extracted error rate describes the physical rather than the logical level, it provides a reproducible metric for characterizing Clifford gate performance under the assumed depolarizing noise model, allowing meaningful comparisons across implementations. Additionally, it yields a value for the fidelity of the transversal logical Clifford gate in the ideal scenario in the absence of state preparation errors and errors in a subsequent QEC cycle. With a fidelity of $\textbf{F=0.9993}$, this corresponds to an error rate approximately four times lower than the effective physical error rate. However, this estimate is based on an extrapolation to error rates substantially lower than in our state preparation, and should therefore be interpreted carefully. Note that this result differs from the simplified theoretical estimate $1 - 21 \cdot p^2$ = 0.9998. 

\begin{figure*}[ht]
    \centering
    \includegraphics[width=0.6\textwidth]{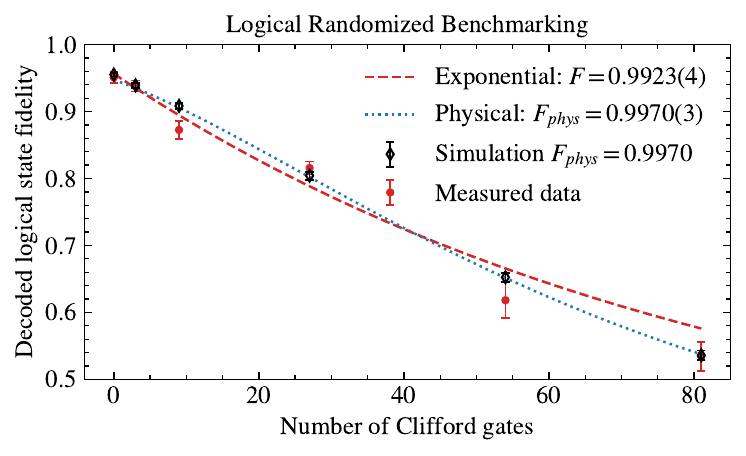}
    \caption{Data and simulation of logical randomized benchmarking. The red points correspond to experimental data with uncertainties, the dashed red line is an exponential (Markovian) fit to them. The blue line represents a fit to a model of physical rather than logical depolarizing noise. The black diamonds correspond to simulations of the experimental quantum circuit including depolarizing physical errors. The simulation results agree with the physical noise model fit and are consistent with the experimental data within error bars. Note that for better visibility the y-axis starts at 0.5. }
    \label{fig:rb_simulation}
\end{figure*}

In order to test the accuracy and relevance of this model, we compare it to simulations of the measured quantum circuit. We use the depolarizing error model described by Eq.~\eqref{eq:error-on-single-qubit-gates} on the Clifford gates and perform a virtual error-correction round in a final step.
Fig.~\ref{fig:rb_simulation} shows the simulated experiment as black diamonds in addition to the experimental data from the main text (red circles). One can see that the circuit simulation and the fitted depolarizing error model Eq.~\eqref{eq:fit-error-RBexp} closely agree. At the same time, they match the experimental data. We observe a modified decay introduced by the error correction scheme compared to the exponential decay: Starting from approximately the same initial value, the simulations lie above the exponential decay fit at around 20 logical Cliffords, before dropping below it after around 40 gates.

\section{Logical State Tomography}
\label{sec:tomogrphy}

In order to characterize an arbitrary qutrit state, and in particular the $\ket{\text{T}}$-state needed for magic state injection, we can perform logical qutrit state tomography. In the qutrit case, this requires characterization of 8 degrees of freedom, while qubit tomography only requires three. Since each qutrit measurement provides two independent probabilities, only four different basis measurements would be necessary if the right bases are chosen. This can be achieved by working with a so-called mutually unbiased bases or MUB, a set of bases in which no two basis vectors are correlated. In our case, we choose to work with:
\begin{widetext}
\begin{equation} 
\bf B_0 = 
{\renewcommand{\arraystretch}{1.8}
\begin{pmatrix}
    \: 1 \quad & \quad 0 \quad &\quad 0  \\
    \: 0 &\quad 1 \quad&\quad 0  \\
    \: 0 & \quad 0 \quad & \quad 1 \quad
\end{pmatrix}, \:
\bf B_1 = 
\begin{pmatrix}
    \frac{1}{\sqrt{3}} & \frac{1}{\sqrt{3}} & \frac{1}{\sqrt{3}} \\
    \frac{1}{\sqrt{3}} & \frac{e^{\frac{-2 i\pi}{3}}}{\sqrt{3}} & \frac{e^{\frac{2 i\pi}{3}}}{\sqrt{3}} \\
    \frac{1}{\sqrt{3}} & \frac{e^{\frac{-2 i\pi}{3}}}{\sqrt{3}} & \frac{e^{\frac{-2 i\pi}{3}}}{\sqrt{3}}
\end{pmatrix}, \:
\bf B_2 = 
%{\renewcommand{\arraystretch}{1.8}
\begin{pmatrix}
    \frac{1}{\sqrt{3}} & \frac{1}{\sqrt{3}} & \frac{1}{\sqrt{3}} \\
    \frac{1}{\sqrt{3}} & \frac{e^{\frac{-2 i\pi}{3}}}{\sqrt{3}} & \frac{e^{\frac{2 i\pi}{3}}}{\sqrt{3}} \\
    \frac{e^{\frac{2 i\pi}{3}}}{\sqrt{3}} & \frac{e^{\frac{-2 i\pi}{3}}}{\sqrt{3}} & \frac{1}{\sqrt{3}}
\end{pmatrix}, \:
\bf B_3 = 
\begin{pmatrix}
    \frac{1}{\sqrt{3}} & \frac{1}{\sqrt{3}} &\frac{1}{\sqrt{3}} \\
    \frac{1}{\sqrt{3}} & \frac{e^{\frac{2 i\pi}{3}}}{\sqrt{3}} & \frac{e^{\frac{-2 i\pi}{3}}}{\sqrt{3}} \\
    \frac{e^{\frac{-2 i\pi}{3}}}{\sqrt{3}} & \frac{e^{\frac{2 i\pi}{3}}}{\sqrt{3}} & \frac{1}{\sqrt{3}}
\end{pmatrix}} \nonumber
\end{equation}
\end{widetext}

More precisely, $B_i$ is the basis transformation matrix from the computational basis into the measurement basis of the tomography step. All of these matrices are Clifford operators and can consequently be implemented transversally on the logical qutrit. A density matrix is reconstructed via linear inversion to obtain an unbiased estimate for our main quantity of interest, the fidelity.

\section{Fault-tolerance of the magic state preparation}
\label{sec:ft-state-prep}

\begin{figure*}[hb]
    \centering
    \includegraphics[width=0.68\textwidth]{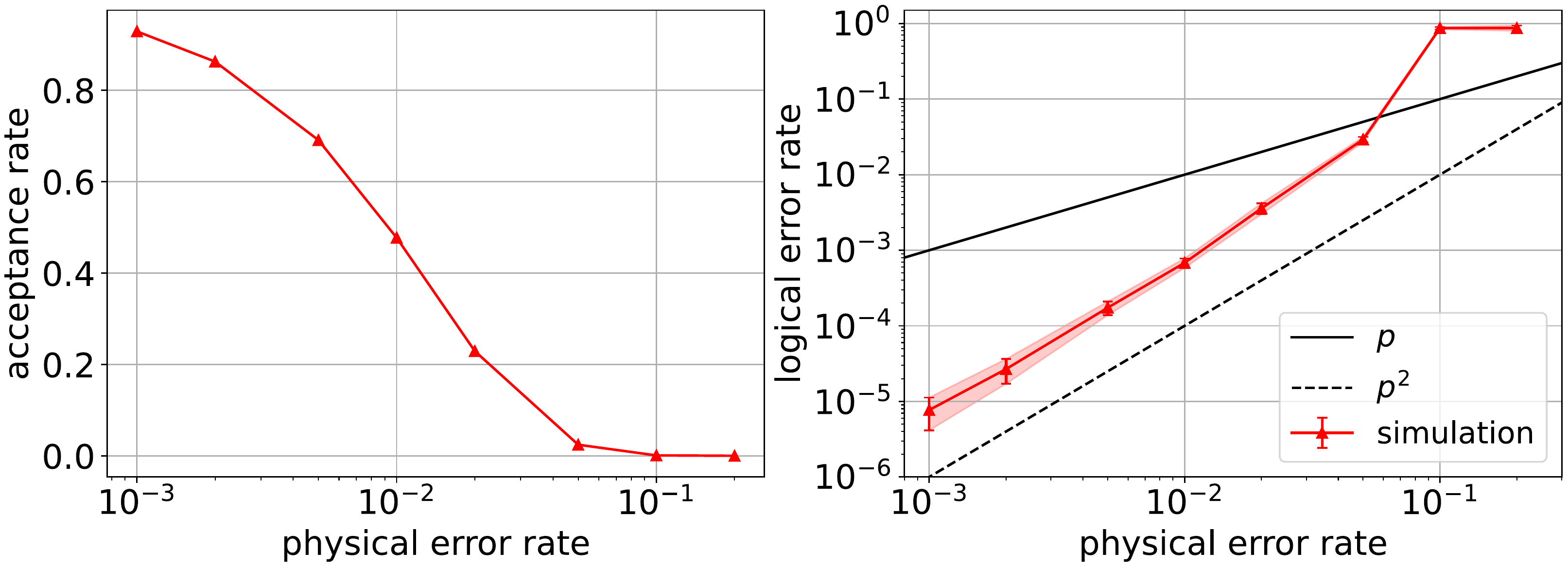}
    \caption{Simulation result for the fault-tolerant $\ket{T}_{\rm L}$-state encoding. Acceptance rate and logical error rate as a function of the physical error rate.}
    \label{fig:T-state-simulations}
\end{figure*}

To showcase that the $T_L$-state preparation circuit in Fig.~4 in the main text is indeed fault-tolerant, we perform state-vector simulations of the encoding circuit using circuit level depolarizing noise [Eq.~\eqref{eq:error-on-single-qubit-gates}]. A virtual round of error correction is then applied, followed by a calculation of the state overlap between the corrected state and the ideal state. Fig.~\ref{fig:T-state-simulations} shows the acceptance rate (left) and the logical error rate (right) as a function of the physical error rate. We observe the expected quadratic scaling for a fault-tolerant implementation.

\end{document}